\documentclass[fleqn,usenatbib,useAMS]{mnras}

\usepackage{graphicx}	
\usepackage{amsmath}	
\usepackage{amssymb}	
\usepackage{multicol}        
\usepackage{bm}		
\usepackage{pdflscape}	

\usepackage{float}
\usepackage{placeins}
\usepackage[normalem]{ulem}

\newcommand{\kms}{\,km\,s$^{-1}$} 
\newcommand{\sigmav}{\sigma_{V}}
\newcommand{\mach}{\mathcal{M}}
\newcommand{\alphavir}{\alpha_\mathrm{vir}}
\newcommand{\rhoSFR}{\rho_\mathrm{SFR}}
\newcommand{\SFRff}{\mbox{\rm SFR}_{\rm ff}}
\newcommand{\eq}[1]{Eq.~\ref{#1}}
\newcommand{\cc}{\mbox{ cm}^{-3}}
\newcommand{\ergs}{\mbox{ erg s}^{-1}}
\newcommand{\nmean}{n_\mathrm{mean}}

\newcommand{\comment}[1]{}

\usepackage[T1]{fontenc}
\usepackage{ae,aecompl}

\usepackage{newtxtext,newtxmath}

\usepackage{xcolor}

\title[Impact of relativistic jets on star formation]{Impact of relativistic jets on the star formation rate: a turbulence-regulated framework}

\author[Mandal et al.]{Ankush Mandal$^{1}$,\thanks{Contact e-mail: \href{mailto:ankushm@iucaa.in}{ankushm@iucaa.in}} Dipanjan Mukherjee$^{1}$,\thanks{Contact e-mail: \href{mailto:dipanjan@iucaa.in}{dipanjan@iucaa.in}} Christoph Federrath$^{2,3}$, Nicole P.~H.~Nesvadba$^{4}$, \newauthor Geoffrey V.~Bicknell$^{2}$, Alexander Y.~Wagner$^{5}$, Moun Meenakshi$^{1}$\\
$^{1}$Inter-University Centre for Astronomy and Astrophysics, Pune-411007, India\\
$^{2}$Research School of Astronomy and Astrophysics, Australian National University, Canberra, ACT 2611, Australia\\
$^{3}$ARC Centre of Excellence for Astronomy in Three Dimensions (ASTRO-3D), Canberra ACT 2601, Australia\\
$^{4}$Universit\'e de la C\^ote d’Azur, Observatoire de la C\^ote d’Azur, CNRS, Laboratoire Lagrange, Bd de l’Observatoire, CS 34229,06304 Nice cedex 4, France\\
$^{5}$University of Tsukuba, Center for Computational Sciences, Tennodai 1-1-1, 305-0006, Tsukuba, Ibaraki, Japan}

\date{}

\pubyear{2021}

\begin{document}
\label{firstpage}
\pagerange{\pageref{firstpage}--\pageref{lastpage}}
\maketitle

\begin{abstract}
We apply a turbulence-regulated model of star formation to calculate the star formation rate (SFR) of dense star-forming clouds in simulations of jet-ISM interactions. The method isolates individual clumps and accounts for the impact of virial parameter and Mach number of the clumps on the star formation activity. This improves upon other estimates of the SFR in simulations of jet--ISM interactions, which are often solely based on local gas density, neglecting the impact of turbulence. We apply this framework to the results of a suite of jet-ISM interaction simulations to study how the jet regulates the SFR both globally and on the scale of individual star-forming clouds. We find that the jet strongly affects the multi-phase ISM in the galaxy, inducing turbulence and increasing the velocity dispersion within the clouds. This causes a global reduction in the SFR compared to a simulation without a jet. The shocks driven into clouds by the jet also compress the gas to higher densities, resulting in local enhancements of the SFR. However, the velocity dispersion in such clouds is also comparably high, which results in a lower SFR than would be observed in galaxies with similar gas mass surface densities and without powerful radio jets. We thus show that both local negative and positive jet feedback can occur in a single system during a single jet event, and that the star-formation rate in the ISM varies in a complicated manner that depends on the strength of the jet-ISM coupling and the jet break-out time-scale.
\end{abstract}

\begin{keywords}
methods: numerical -- galaxies: jets -- galaxies: ISM -- galaxies: star formation -- Galaxy: evolution
\end{keywords}



\begingroup
\let\clearpage\relax
\endgroup
\newpage

\section{Introduction}\label{Intro}
For the past few decades, an extensive number of studies has established that outflows from active galactic nuclei (AGN) have a profound effect on the overall evolution of their host galaxy and the formation of stars \citep{Silk_1998,Bower_2006,Croton_2006,Alexander_2012,Fabian_2012}. The feedback from the central black hole is thought to affect the galaxy's evolution via two major pathways. Large-scale outflows are expected to heat the circum-galactic environment, preventing catastrophic cooling and regulating the gas in-fall rate and star formation. This has been explored in detail in several simulations of jet-induced heating of the intracluster medium \citep[such as][]{Gaspari_2011a,Yang_2016,Weinberger_2017,Prasad_2020,Bourne_2021}, and also supported by observational studies \citep{Birzan_2004,Fabian_2012,Morganti_2013}. On the other hand, the local input of the energy and momentum from the outflows into the interstellar medium of the host galaxy is also believed to affect the properties of the entire galaxy \citep{Nesvadba_2007,Nesvadba_2010,Nesvadba_2011,Harrison_2014,Guillard_2015,Alatalo_2015,Bae_2017,Rupke_2017,Wylezalek_2020}. However, how these AGN outflows affect the overall galactic dynamics on different scales, as well as the star formation activity, are still poorly understood \citep{Schawinski_2009,Schawinski_2015}.

Recent observational studies \citep{Ogle_2007,Ogle_2010,Nesvadba_2010,Nesvadba_2011,Alatalo_2014,Alatalo_2015,Lanz_2016,Nesvadba_2021} have shown that the star formation rate (SFR) in some galaxies harbouring radio jets is significantly lower than in standard star-forming galaxies following the Kennicutt-Schmidt relation \citep{Schimdt1959,Schimdt1963,KS_1998a,KS_1998b}. In the early phase of their evolution, radio jets can strongly couple with the host's interstellar medium while emerging from the galactic scales, launching outflows and inducing turbulence and shock heating, as seen in several galaxies \citep[e.g.][]{Nesvadba_2008,Nesvadba_2011,Collet_2016,Murthy_2019,Zovaro_2019,Zovaro_2019b,Venturi_2021}. This has also been found in well-resolved simulations of jet--ISM interactions  \citep[such as,][]{Sutherland_2007,Wagner_2011,Wagner_2012,Mukherjee_2016,Mukherjee_2017,Mukherjee_2018,Mukherjee_2018a}. Thus, an idea is emerging where radio jets may play a major role in transferring power from the AGN to the multiphase ISM and in turn regulate the star formation efficiency in the dense gas. Observations of radio-loud AGN \citep{Morganti_2005,Best_2005,Fu_2009} and several hydrodynamic simulations \citep{Krause_2005,Sutherland_2007,Antonuccio_2008,Mukherjee_2016,Mukherjee_2017} have shown that radio-loud AGNs inject a few percent of their mechanical energy into the ambient gas, causing significant outflows of hot gas. However, these outflows are too weak to completely expel more than a few percent of the total molecular gas from the galaxy. Most of the remaining molecular gas is very inefficient in forming stars \citep{Nesvadba_2011,Alatalo_2015}. Thus, it is still unclear by which mechanism star formation is suppressed in galaxies.

On the other hand, it is not obvious why jet feedback should always be negative. Jets have also been posited to trigger star formation inside the host galaxy \citep{Silk_2005} due to compression by ensuing shocks. Direct evidence of this phenomenon has been found in observed sources, such as Minkowski's object \citep{Croft_2006,Salome_2015,Lacy_2017,Zovaro_2020}, 3C~285 \citep{Salome_2015}, Centaurus~A \citep{Mould_2000,Morganti_2010,Salome_2017}, 4C~41.17 \citep{Dey_1997,Bicknell_2000,Nesvadba_2020}, PKS~2250-41 \citep{Inskip_2008}, and indirect evidence of enhanced SFR in radio-loud AGNs \citep{Zinn_2013,Kalfountzou_2014}. Circumstantial evidence of a positive correlation between star formation activity and the existence of radio jets is plentiful, such as the observation of a large fraction of cold molecular gas in radio galaxies \citep{Emonts_2011}, detection of late-time star formation activity in compact radio-loud AGNs \citep{Dicken_2012,Kalfountzou_2017}, alignment of CO emission along the radio jet \citep{Klamer_2004}, and the existence of a young stellar population in radio galaxies \citep{Aretxaga_2001,Tadhunter_2002,Wills_2002,Baldi_2008,Tadhunter_2011,Rocca_2013}. Recent theoretical studies have also shown that positive feedback is a viable mechanism in which the expanding bow shock from the jet compresses the dense pockets of the clumpy ISM to a high density that cool efficiently, generating potential star-forming sites \citep{Fragile_2004,Silk_2009,gaibler_2012,Fragile_2017,Zubovas_2017,Mukherjee_2018}. Thus, the way in which the jet affects star formation in the host galaxy is still not well understood and requires better modelling of the star formation process. The coexistence of both negative and positive feedback in a single system \citep{Cresci_2015,Shin_2019} complicates the situation further. 
 
From the theoretical point of view, the major challenge in simulating the complex mechanisms of star formation in large-scale simulations stems from the fact that the star formation activity involves physical processes spanning several orders of magnitude in spatial scales (from several tens of kpc to less than pc). This makes it difficult to follow individual collapsing structures while simultaneously keeping track of the global evolution \citep[see][for a recent review]{Vogelsberger_2020}. However, the most significant advance on this issue has been through the implementation of subgrid models of star formation \citep{Cen_1992,Katz_1992,Springel_2003,Li_2017} and stellar feedback \citep{Stingson_2006,Agertz_2013,Vogelsberger_2013,Hopkins_2014}. Nonetheless, these subgrid models rely on fine-tuning different model parameters to reproduce the observational results, such as the Kennicutt-Schmidt relation, the galaxy luminosity function, etc. Moreover, it remains unclear whether these models are able to capture the physical processes on sub-resolution scales successfully, such as the variation in the observed correlation between star formation surface density and gas surface density in individual molecular clouds \citep{Evans_2009,Lada_2010,Heiderman_2010,Guthermuth_2011,Krumholz_2012,Federrath_2013b,Evans_2014,Salim_2015}. The model results differ from extragalactic observations \citep{KS_1998b,Gao_2004,Kennicutt_2007,Genzel_2010}.

Although a significant number of simulations \citep{Cen_1992,Katz_1992,Springel_2003,Agertz_2013,Hopkins_2014,Li_2017} have explored subgrid prescriptions of the underlying small-scale physical processes of star formation, most AGN feedback simulations that require relativistic jet dynamics still lack this. Simple models of estimating star formation in the context of AGN feedback simulations \citep{gaibler_2012,Zubovas_2014,Bieri_2016} consist of creating sink particles that represent a star cluster where the density is above some user-defined threshold value, and the efficiency of star formation is set to a constant value irrespective of other physical conditions (notably turbulence) in that region. However, the studies of star formation in individual molecular clouds have established that the efficiency of star formation inside a cloud is significantly regulated by different physical parameters, such as the velocity dispersion, gravitational binding energy, temperature, magnetic field, etc. \citep[e.g., see the reviews by][]{Mac_Low_2004,McKee_2007}. Nevertheless, the implementation of small-scale processes in large-scale simulations is a challenging task. However, galactic scale ($\sim$ few kpc) simulations \citep{Sutherland_2007,Wagner_2011,Wagner_2012,Mukherjee_2016,Bieri_2016,Mukherjee_2018a,Mukherjee_2018,Cielo_2018} or zoom-in cosmological simulation of isolated galaxies \citep{Wetzel_2016,Li_2017,Hopkins_2018,Wheeler_2019,Agertz_2020} allow us to resolve individual star-forming clouds (few tens of pc) to some extent. Thus, statistically considering the details of the star formation mechanism from small-scale studies \citep{Krumholz_2005,Federrath_2012} in intermediate-scale AGN feedback simulations promises to give a better understanding of star formation and the effects of feedback.

In this paper, our primary motivation is to estimate the impact of relativistic jets on the star formation rate (SFR) and its evolution in the simulations presented in \citet{Mukherjee_2018}, by identifying potential star-forming regions using a modified CLUMPFIND module (see Appendix~\ref{append:Fellwalker} for details) based on the FellWalker algorithm \citep{BERRY_2015}. We model the star formation mechanism following the semi-analytical method proposed by \citet{Krumholz_2005} and \citet{Federrath_2012} from their study of individual molecular clouds. We study the variation of different cloud properties such as the velocity dispersion, virial parameter (ratio between the kinetic energy and potential energy), etc., which regulate star formation on small scales. 

The paper is structured as follows. In Sec.~\ref{methods} we briefly describe the method of estimating the SFR inside a molecular cloud. In Sec.~\ref{result} we present our simulation results. We discuss the implications of our results in Sec.~\ref{discussion}. Finally, in Sec.~\ref{conclusion}, we summarise and conclude.

\section{Methods}
\label{methods}
\subsection{Turbulence-regulated star formation}\label{background}
Stars form in dense, cold, turbulent molecular clouds on scales where the kinetic, magnetic, and thermal energy of the gas cannot prevent  local gravitational collapse. There have been extensive studies about the different physical mechanisms affecting the process of star formation \citep{Silk_1997,Tan_2000,Kravtsov_2003,Tassis_2004,Li_2005,Padoan_2014,Krumholz_2019}. Turbulence-regulated star formation is one theory \citep{Krumholz_2005,Padoan_2011,Hennebelle_2011,Federrath_2012}, which suggests that turbulent motions prevent (or at least slow down) global gravitational collapse on the cloud-scale \citep[see reviews by][]{Mac_Low_2004,Elmegreen_2004,McKee_2007}. However, on smaller scales where the gravitational energy exceeds the turbulent kinetic, magnetic, and thermal energy, dense cores can eventually collapse to form stars.

Various semi-analytical models of turbulence-regulated star formation suggest that the collapse occurs approximately at the sonic scale \citep{Federrath_2021}, where the turbulent velocity dispersion is of the order of the thermal sound speed \citep{Krumholz_2005,Federrath_2012}. Below the sonic scale, gravity dominates over any other form of energy (e.g., turbulence, magnetic fields, thermal pressure) that would oppose the collapse. Thus, if the Jeans length (the critical length for collapse) becomes smaller than the local sonic length, the region will eventually collapse. This translates to a critical density ($\rho_\mathrm{crit}$) for star formation of a region. Thus, the theoretical estimation of the SFR in a molecular cloud starts by determining the mass fraction above the critical density. 

It is a well established result that in supersonic, isothermal turbulence, the density PDF approximately follows a log-normal distribution \citep{Vazquez_1994,passot_1998,Federrath_2010,Federrath_Ban_2015,Kritsuk_2017}. However, even if the gas is not isothermal, when cooled to $T \lesssim 100\,\mathrm{K}$, the log-normal approximation of the density PDF still remains valid \citep{Kortgen_2019,Mandal_2020}. In terms of the logarithmic density $s=\ln (\rho/\rho_0)$ (where $\rho_0$ is the mean density of the cloud), the density PDF is expressed as,
\begin{equation}\label{eq:density PDF}
    p_s(s)=\frac{1}{\sqrt{2\pi\sigma_s^2}}\exp\left(-\frac{(s-s_0)^2}{2\sigma_s^2}\right).
\end{equation}
Here $s_0$ is the mean of the lognormal distribution and $\sigma_s$ is the dispersion of the logarithmic density fluctuation. Extensive theoretical and numerical studies \citep{Padoan_1997a,passot_1998,Federrath_2008,Price_2011,Konstandin_2012,Molina_2012,Nolan_2015,Beattie_2021} have shown that the logarithmic density dispersion ($\sigma_s$) for hydrodynamic turbulence depends on the turbulent Mach number ($\mach=\sigmav/c_s$; where $\sigmav$ and $c_s$ are the 3D velocity dispersion and sound speed respectively) as,
\begin{equation}\label{eq:sigma_s}
    \sigma_s^2=\ln\left(1+b^2\mach^2\right).
\end{equation}
Here $b$ is the turbulence driving parameter, which represents the ratio of power in compressive modes to solenoidal modes \citep{Federrath_2008,Federrath_2010} and has a value between 1/3 (purely solenoidal) and 1 (purely compressive), respectively. Here we assume a value of $b=0.5$, corresponding to a common mixed mode of driving of the turbulence. However, variations do exist between different regions and clouds \citep{Federrath_2016,Menon_2020,Menon_2021}.

From the density PDF of the star-forming clouds, one can readily find the mass fraction that will form stars per global, average free-fall time (defined as $\SFRff$) as \citep{Federrath_2012},
\begin{align}\label{eq:SFRff}
    \SFRff &= \frac{\epsilon}{\phi_t}\int_{s_\mathrm{crit}}^\infty \frac{t_\mathrm{ff}(\rho_0)}{t_\mathrm{ff}(\rho)}\frac{\rho}{\rho_0} p(s)ds \nonumber \\
    &= \frac{\epsilon}{2\phi_t}\left[1+\mathrm{erf}\left(\frac{\sigma_s^2-s_\mathrm{crit}}{\sqrt{2\sigma_s^2}}\right)\right]\exp\left(\frac{3}{8}\sigma_s^2\right),
\end{align}
where the density-dependent freefall time is evaluated inside the integral, which makes this a multi-freefall model \citep{Hennebelle_2011} as opposed to the single-freefall model of \citet{Krumholz_2005}. The parameter $\phi_t$ is a numerical factor in the range $0.3-2$ \citep{Krumholz_2005,Federrath_2012}. The parameter $\epsilon$ is the efficiency of converting a given gas mass to stars, represented as a fraction. For small-scale studies that resolve individual star-forming cores inside a molecular cloud \citep{Matzner_2000,Krumholz_2005,Alves_2007,Andre_2010,Federrath_2012,Federrath_2014}, $\epsilon$ represents the fraction of the core mass that eventually ends up in stars. It is usually found to lie in the range $0.3-0.7$. However, in this study, the parameter $\epsilon$ is related to the fraction of the global gas mass of a molecular cloud that may eventually turn into stars. We choose $\epsilon$ to be a free parameter, which we calibrate based on the observational Kennicutt-Schmidt relation \citep{KS_1998a}, as discussed in Sec.~\ref{no jet simulation}.

\subsection{Estimation of SFR in the simulations}\label{impletementation}
We use the above formalism to estimate the SFR at a given time snapshot of a simulation in post-processing. From Eq.~\eqref{eq:SFRff} we see that the key ingredients determining the SFR are the relative strength of the gravitational potential energy, the velocity dispersion, and the Mach number of the local dense gas-rich regions of a galaxy. Hence, to estimate the SFR in the multi-phase gas distributions in our simulations, at a given time step, we first identify the potential star-forming regions by finding dense gas clumps using a clump-finding algorithm. Subsequently, we compute the properties of each clump, such as its potential energy, velocity dispersion, and Mach number, required to evaluate the expected SFR based on Eq.~\eqref{eq:SFRff}. 

We briefly summarize the steps in computing SFR below.
\begin{enumerate}    
    \item First, we identify dense contiguous gas clumps using the FellWalker algorithm \citep{BERRY_2015}, which uses a gradient tracing scheme. The clumps are identified for regions with density greater than a threshold of $n = 100\,\mathrm{cm^{-3}}$. Clumps with a volume less than a minimum number of computational cells (here $(10^3)$) are excluded. This ensures that there is enough resolution to compute the local statistical quantities inside a given clump, such as the velocity dispersion and Mach number. A brief summary of the method and the assumptions of various parameters used in this work are in Appendix~\ref{append:Fellwalker}.    

    \item We compute the gravitational potential ($\Phi(\boldsymbol{r})$) due to the gas mass inside each clump by extracting the given clump from the simulation domain separately and solving the 3D Poisson equation using a successive over-relaxation (SOR) scheme with Chebyshev acceleration \citep{Press_1992}. The boundary conditions were evaluated using a multipole expansion of the density distribution, which assumes the potential at $r\to\infty$ is 0. The procedure for solving the Poisson equation is further elaborated in Appendix~\ref{append:Poisson}.

    \item For each dense clump in a simulation we calculate the following parameters in the centre-of-mass frame of the cloud:
    \begin{enumerate}
        \item The mass-weighted 3D velocity dispersion ($\sigmav$) \footnote{The mass-weighted 3D velocity is defined as $\sigmav=\sqrt{\sigma_{x_1}^2+\sigma_{x_2}^2+\sigma_{x_3}^2}$. Here, $\sigma_{x_m}$ is the mass-weighted velocity dispersion along $m^\mathrm{th}$ dimension (e.g., $x$, $y$ and $z$) given by
        \begin{equation*}
            \sigma_{x_m}^2=\frac{\sum_{i,j,k}M(i,j,k)[v_{x_m}(i,j,k)-\Bar{v}_{x_m}]^2}{\sum_{i,j,k}M(i,j,k)},
        \end{equation*}
        where $v_{x_m}$ is the centre-of-mass velocity of the cloud along the $x_m$ axis, given by
        \begin{equation*}
            \Bar{v}_{x_m}=\frac{\sum_{i,j,k}M(i,j,k)v_{x_m}(i,j,k)}{\sum_{i,j,k}M(i,j,k)}.
        \end{equation*}
        Here, $M(i,j,k)$ and $v_{x_m}(i,j,k)$ are the mass and $x_m$ component of the velocity of the cell with index $(i,j,k)$, respectively.
        }.
        \item Total kinetic ($E_\mathrm{kin}$), self gravitational potential energy ($E_\mathrm{grav}=\sum_{ijk}\Phi_{ijk}M_{ijk}$, where $M_{ijk}$ is the mass of the (i,j,k) cell) and thermal ($E_\mathrm{th}$) energy, and subsequently the virial parameter $\alpha_\mathrm{vir}=2E_\mathrm{kin}/E_\mathrm{grav}$.
        \item The mass-weighted rms sound speed ($c_s$) and the Mach number $\mathcal{M}=\sigmav/c_s$.
    \end{enumerate}
   \item Assuming a log-normal density PDF for each cloud, we calculate the density dispersion ($\sigma_s$) using Eq.~\eqref{eq:sigma_s}. Further, the comparison between the Jeans length to the sonic length gives the critical density ($\rho_\mathrm{crit}$) for collapse, which is calculated by (see Appendix~\ref{append:SFR theory} for details)
    \begin{equation}\label{rho_critical}
        \rho_\mathrm{crit}=\rho_0\left[\left(\frac{\pi^2}{5}\right)\phi_x^2\alpha_\mathrm{vir}\mathcal{M}^2\right],
    \end{equation}
    with $\phi_x$ a numerical parameter of order unity calibrated in \citet{Krumholz_2005} and \citet{Federrath_2012}. We then estimate the SFR per free-fall time ($\SFRff$) by evaluating Eq.~\eqref{eq:SFRff} for each cloud.
    
    \item Lastly, the SFR of a cloud of mass $M_c$ is computed using \citep{Federrath_2012},
    \begin{equation}\label{eq:SFR}
        \mathrm{SFR}=\SFRff\frac{M_c}{t_\mathrm{ff}(\rho_0)}.
    \end{equation}
    Here, $t_\mathrm{ff}(\rho_0)$ is the freefall time at the mean density ($\rho_0$) of the cloud.
\end{enumerate}

\section{Results}
\label{result}
We have applied the method outlined above to compute the SFR of four simulations presented in \citet{Mukherjee_2018}. The simulations explore the evolution of a relativistic jet of power $P_{\rm jet} = 10^{45} \ergs$ interacting with an inhomogeneous gas disc. Table~\ref{simulation list} provides the list of simulations that have been considered in this work. The nomenclature is the same as that used in \citet{Mukherjee_2018}. Three of the simulations have different angles between the jet launch axis and the normal to the disc, and a fourth is a simulation without a jet (referred to as `no-jet' in Table~\ref{simulation list}), which serves as the control case with which to compare the evolution of the SFR in simulations that have jets. We refer the reader to \citet{Mukherjee_2018} for further details of the simulation setup and the evolution of the dynamics of the gas disc impacted by the jet. The results of the estimates of the SFR of the above simulations are discussed in the following sections.

\begin{table}
 \caption{List of simulations from \citet{Mukherjee_2018} used in this study.}
 \label{simulation list}
 \begin{tabular*}{\columnwidth}{c @{\extracolsep{\fill}} cccccc}
  \hline
  Simulation & Jet power & $n_{w0}$ & $\theta_\mathrm{inc}$ $^b$ & $\Gamma$ $^c$ & Gas mass\\
  label & $(\mathrm{erg\,s^{-1}})$ & $(\mathrm{cm^{-3}})$ $^a$ & & & $(10^9\,\mathrm{M_\odot})$\\
  \hline
  no-jet & - & 200 & - & - & 5.71 \\
  B & $10^{45}$ & 200 & $0^\circ$  & 5 & 5.71 \\
  D & $10^{45}$ & 200 & $45^\circ$  & 5 & 5.71 \\
  E & $10^{45}$ & 200 & $70^\circ$  & 5 & 5.71 \\
  \hline
 \end{tabular*}
 \footnotesize $^a$ Gas density at the centre of the disc. \\
 \footnotesize $^b$ Angle of inclination of the jet with respect to the normal to the disc.\\
 \footnotesize $^c$ Lorentz factor of the jet plasma at the launch time.
\end{table}

\subsection{`No-jet' Simulation (Control)}\label{no jet simulation}
\subsubsection{Calibrating the SFR efficiency}
A widely used practice for estimating the SFR in large-scale simulations \citep{Springel_2003,Schaye_2008,Dubois_2008,gaibler_2012,Bieri_2016} comprises of calibrating some parameters of the model, e.g., the global star formation efficiency per freefall time, the star formation time-scale, etc., so that the galaxy follows the standard Kennicutt-Schmidt relation \citep{KS_1998a,KS_1998b}. We also adopt this method in our study and set the efficiency parameter $\epsilon$ in Eq.~\eqref{eq:SFRff} to ensure that the `no-jet' simulation lies on the Kennicutt-Schmidt (hereafter KS) relation. We find that a value of $\epsilon=0.015$ puts the galaxy exactly on the KS line at $\sim 0.8\,\mathrm{Myr}$. This time was chosen so that the inhomogeneous disc has enough time to settle down after the initialization of the simulation. This sets a reference to compare the effect of jet feedback on the SFR of the galaxy explored in the later sections.

As pointed out earlier, $\epsilon$ in Eq.~\eqref{eq:SFRff} is normally $\sim 0.5$, i.e., the core-to-star formation efficiency. This means about half the gas in a dense core (defined as an object of size $\sim 0.1\,\mathrm{pc}$ and respective density) falls onto the star, while the other half is expelled by proto-stellar jets and outflows \citep[see e.g.,][]{Federrath_2014}. Here, on the other hand, the $\epsilon$ parameter cannot be interpreted as the core-to-star efficiency as we are evaluating the global star formation for a whole molecular cloud, and not the individual star-forming cores. Therefore, the $\epsilon$ here is better described as a cloud-to-star formation efficiency (rather than a core-to-star fraction), whose values are chosen so that the SFR of the galaxy lies on the KS relationship. The values of $\epsilon$ used are 1--2\%, consistent with other recent studies \citep[e.g.][]{Federrath_2013a,Salim_2015}, meaning that typically only a few percent of the gas in a cloud forms stars. Thus, our choice of $\epsilon$ implicitly assumes that regulating processes such as stellar feedback are operating to place the galaxy on the KS relationship even though such processes are not explicitly included in the simulations.

In summary, the value of $\epsilon$ is tuned to give us a typical SFR of a typical galaxy (i.e., following the KS relation) in the absence of a relativistic jet. This calibration therefore provides us with a reference for comparing the effect of the relativistic jet in the subsequent analysis, allowing us to quantify by what relative amount the SFR changes when jet feedback is included.

\begin{figure}
    \centering
    \includegraphics[width=\linewidth]{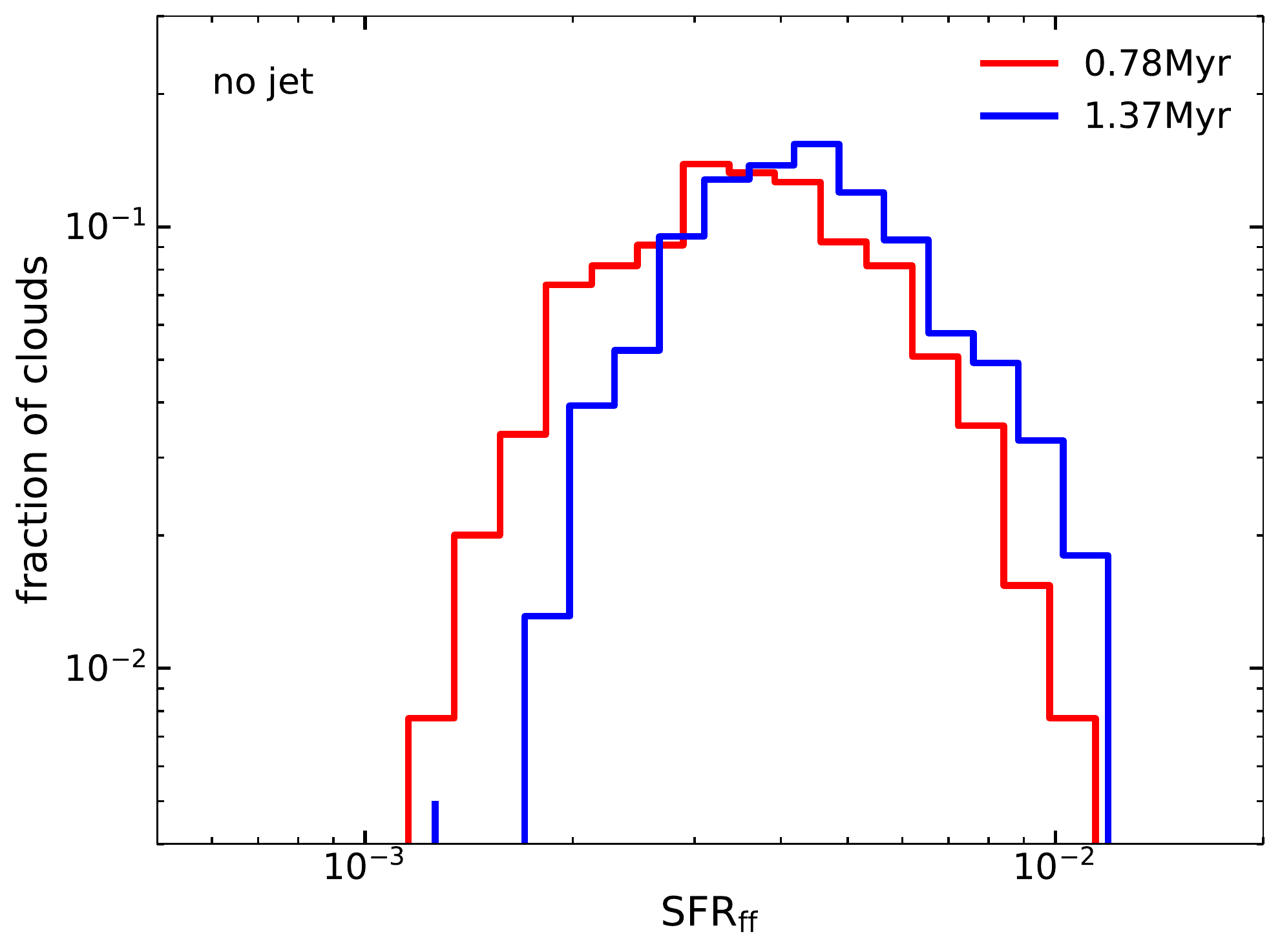}
    \caption{Distribution of $\SFRff$ as the number fraction of the clouds for the `no-jet' simulation. The red and blue solid lines correspond to 0.78 and 1.37 Myr, respectively. We see that the mean value of $\SFRff$ increases slightly at the later time as the gas cools down and the Mach number, and hence the SFR increases.}
    \label{SFRff PDF nojet}
\end{figure}
\subsubsection{Star formation rate per freefall time ($\SFRff$)}\label{SFRff nojet}
Having fixed all the parameters of our star formation model, we explore how this method of estimating the star formation compares with previous theoretical and observational studies. One of the major parameters that characterise the SFR inside a cloud is the SFR per free-fall time ($\SFRff$). In galactic-scale simulations \citep{Springel_2003,Schaye_2008,Dubois_2008,gaibler_2012,Bieri_2016}, this parameter is set to a constant value of (1-10\%). 

However, simulations of individual molecular clouds have shown that $\SFRff$ depends on the physical  properties  and the nature of turbulence intrinsic to the cloud \citep{Krumholz_2005,Padoan_2011,Hennebelle_2011,Federrath_2012}, as discussed earlier in Sec.~\ref{methods}. Typically, the $\SFRff$ can have a wide distribution about a mean of a few percent \citep{Krumholz_2005,Padoan_2011,Federrath_2012,Padoan_2012}. In Fig.~\ref{SFRff PDF nojet}, we present the distribution of $\SFRff$ of the `no-jet' simulation at 0.78 and 1.37~Myr. We notice that the $\SFRff$ has a broad distribution with a mean value of $\sim 0.0004$. Our estimates of $\SFRff$ yields a lower value of $\SFRff$ than what is typically found in the simulations \citep[see][for a review and references therein]{Krumholz_2019}. This likely occurs for three reasons. Firstly, our simulations have only considered atomic cooling with the temperature floor set to $1000\,\mathrm{K}$. This results in relatively lower Mach numbers for the simulated clouds than in real star-forming clouds with much lower temperatures (a few tens of Kelvin) \citep{Gratier_2010,Hughes_2010,Heyer_2015,Jameson_2019}, making them more efficient in forming stars. Secondly, the initialization of the velocity dispersion in the simulations of \citet{Mukherjee_2018} was chosen to be ($\sim 40\,\mathrm{km\,s^{-1}}$), and the resolution of the simulation also restricts the mean density of the clouds, which makes the clouds tenuous. This results in a relatively high virial parameter, which gives a much lower value of $\SFRff$. Thirdly, the $\SFRff$ in small-scale studies \citep[e.g.][]{Krumholz_2005,Federrath_2012} was estimated for individual molecular clouds and not calibrated such that a large kpc-scale gas disc matches the KS relation, as we have done here. As discussed above, requiring an exact match with the KS relation for the given gas surface densities in our simulations implies a lower SFR efficiency $\epsilon$.

Nonetheless, we must emphasize that calibrating $\epsilon$ in \eq{eq:SFRff} such that the kpc-scale gas disc of the `no-jet' simulation matches the KS relation, even if it yields a relatively lower $\SFRff$, gives us a suitable `no-jet' sample to which the SFR of the jetted simulations can be compared.

\begin{figure}
  \centering
  \includegraphics[width=\linewidth]{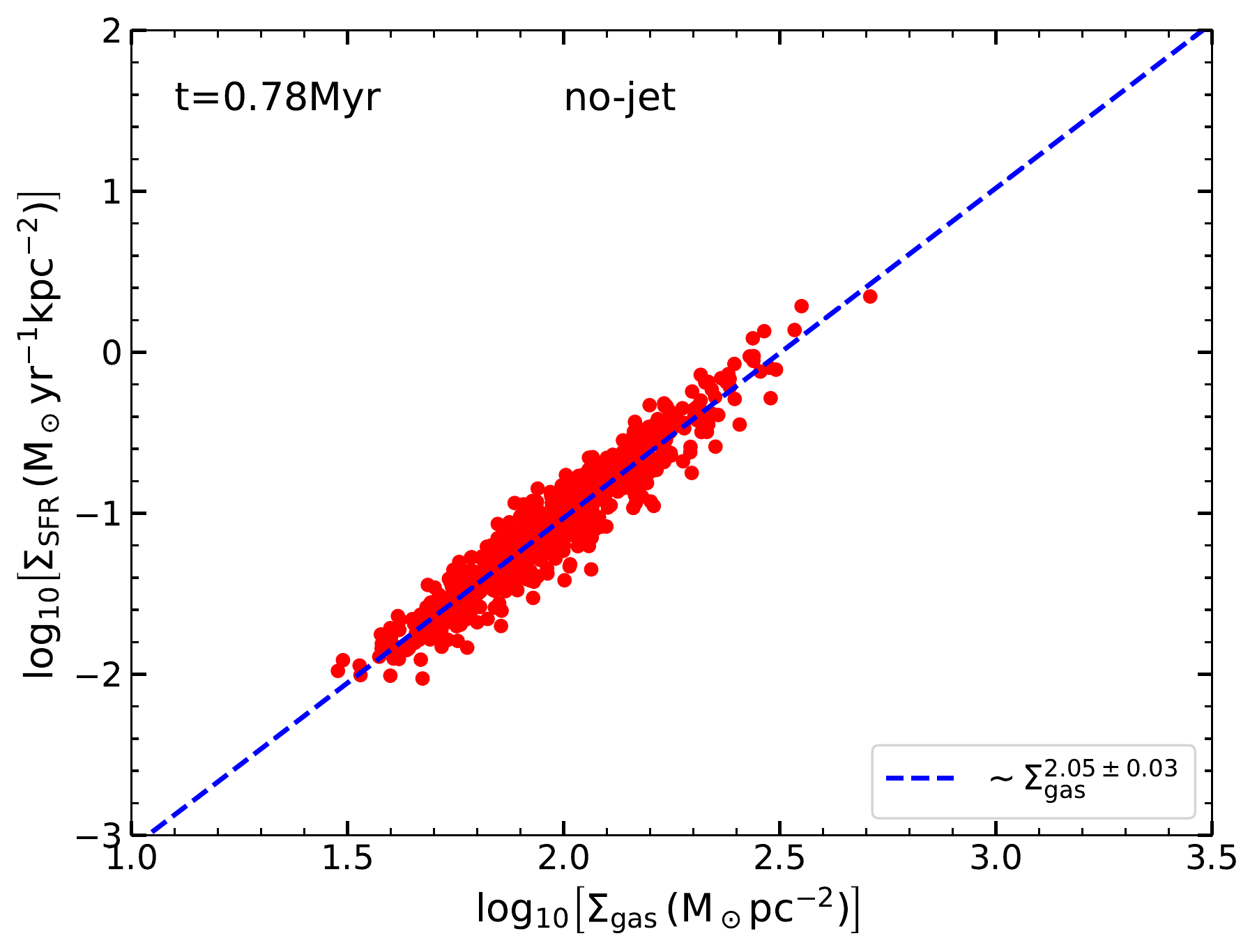}
\caption{SFR surface density ($\Sigma_\mathrm{SFR}$) as a function of gas mass surface density ($\Sigma_\mathrm{gas}$) for individual clouds in the `no-jet' simulation (red circles) at 0.78~Myr. The blue dashed line is a power-law fit to the data.}
\label{Stellar surface density no-jet}
\end{figure}
\subsubsection{SFR surface density vs gas surface density}
We present the SFR surface density $(\Sigma_\mathrm{SFR})$ as a function of gas surface density ($\Sigma_\mathrm{gas}$) for individual clouds in the `no-jet' simulation at 0.78~Myr in Fig.~\ref{Stellar surface density no-jet}. To compute the gas mass surface density and the SFR surface density, we first calculate the area by projecting all clumps on the $x\mbox{-}y$ plane and finding the number of pixels that have a non-zero density value. The mean SFR surface density and gas mass surface density are then evaluated by dividing the global value of the corresponding quantity by the total projected area of the clumps computed earlier. We notice that $\Sigma_\mathrm{gas}$ and $\Sigma_\mathrm{SFR}$ are tightly correlated and are well fitted by a power law (dashed blue line) with an index of $~\sim 2$. Similar results have also been obtained by high-resolution observations \citep[e.g.,][]{Momose_2013,Wilson_2019}, where the mean slope was found to be $\sim 1.8$. This is different from the slope of the $\Sigma_\mathrm{SFR}\mbox{-}\Sigma_\mathrm{gas}$ relation inferred from galaxy-integrated measurement by \citet{KS_1998b}. However, such discrepancies between the inferred correlation on large scales compared to individual molecular cloud scales have been addressed by a number of studies. Several authors \citep[e.g.,][]{Evans_2009,Lada_2010,Heiderman_2010,Guthermuth_2011} have shown that in resolved molecular clouds the correlation between the star formation surface density and the gas mass surface density strongly deviates from the Kennicutt-Schmidt relation. This is likely due to the inclusion of more tenuous, non-star-forming gas in the observations over kpc scales compared to the scales of individual star-forming clouds \citep[see][for a review]{Kennicutt_2012}. We also note that an exponent of $1.5$ can be explained if the SFR is primarily proportional to the gas density divided by the freefall time of the gas \citep[][]{Schimdt1959,Elmegreen_1994,Wong_2002,Krumholz_2007}. However, the incorporation of the impact of local physical quantities such as the virial parameter and Mach number, can change the dependence between the gas mass surface density and the SFR surface density, which has been addressed by several authors \citep[e.g.,][]{Federrath_2013b,Salim_2015}. Thus, our results on the estimates of resolved star-forming clouds are in agreement with the observations on similar scales. 

\subsection{Evolution of the dynamical quantities of an ISM impacted by a jet}\label{Sim D}
In this section, we present the effect of a jet on the star formation inside a gas disc. We consider Sim.~D, with the jet inclined at an angle of $45^\circ$ to the disc plane since it has been run for the longest time compared to the other jet simulations. The inclination of the jet creates a substantial region of interaction with the ISM for a relatively long time, and this is ideal for exploring the impact of the jet on the SFR. In the following sections, we show the evolution of the different dynamical quantities that regulate the SFR.

\begin{figure*}
\centerline{
\def\arraystretch{1.0}
\setlength{\tabcolsep}{0.0pt}
\begin{tabular}{lcr}
  \includegraphics[width=0.5\linewidth]{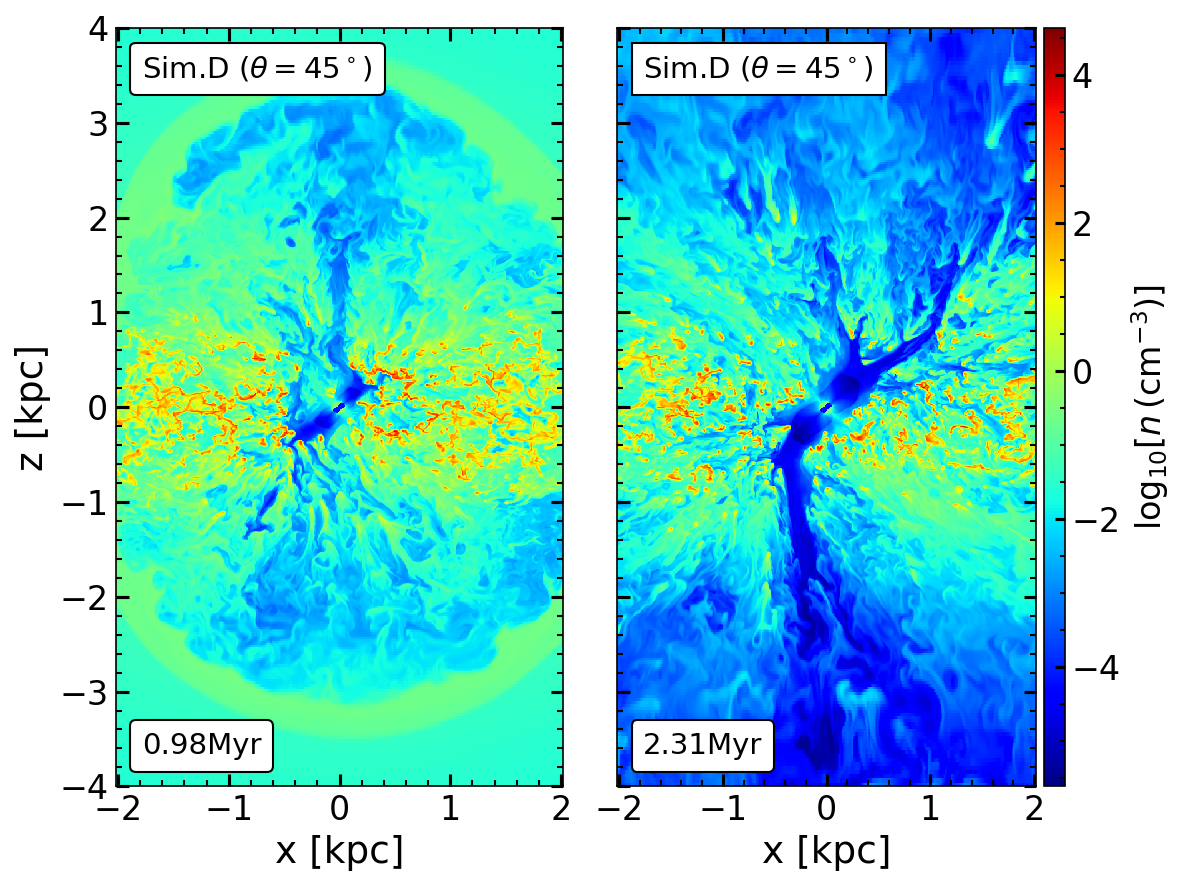} &
  \includegraphics[width=0.5\linewidth]{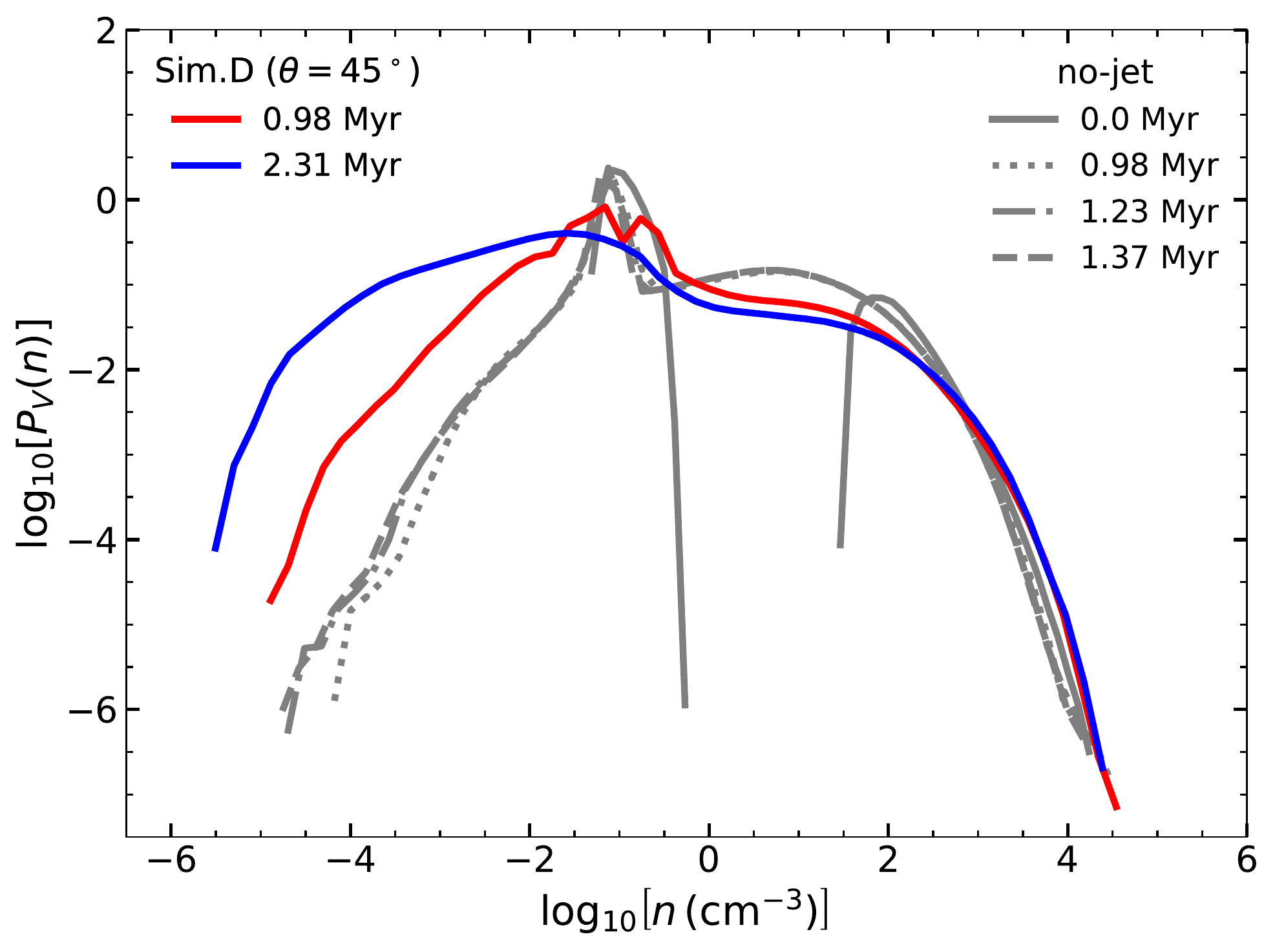}
\end{tabular}}
  \caption{The left and middle panels show the number density distribution in the $x\mbox{-}z$ plane at 0.98 and 2.31~Myr for Sim.~D. At 0.98~Myr, the jet is well inside the disc, but at 2.31~Myr, the jet has already created a channel through the ISM. The right panel shows the volume-weighted probability distribution function (PDF) of the number density. The grey lines correspond to the density PDF for the `no-jet' simulation at different times. The red and blue solid lines represent the volume-weighted density PDF for Sim.~D at 0.98 and 2.31~Myr.}
  \label{density pdf sim D}
\end{figure*}
\begin{figure*}
\centerline{
\def\arraystretch{1.0}
\setlength{\tabcolsep}{0.0pt}
\begin{tabular}{lcr}
  \includegraphics[width=0.5\linewidth]{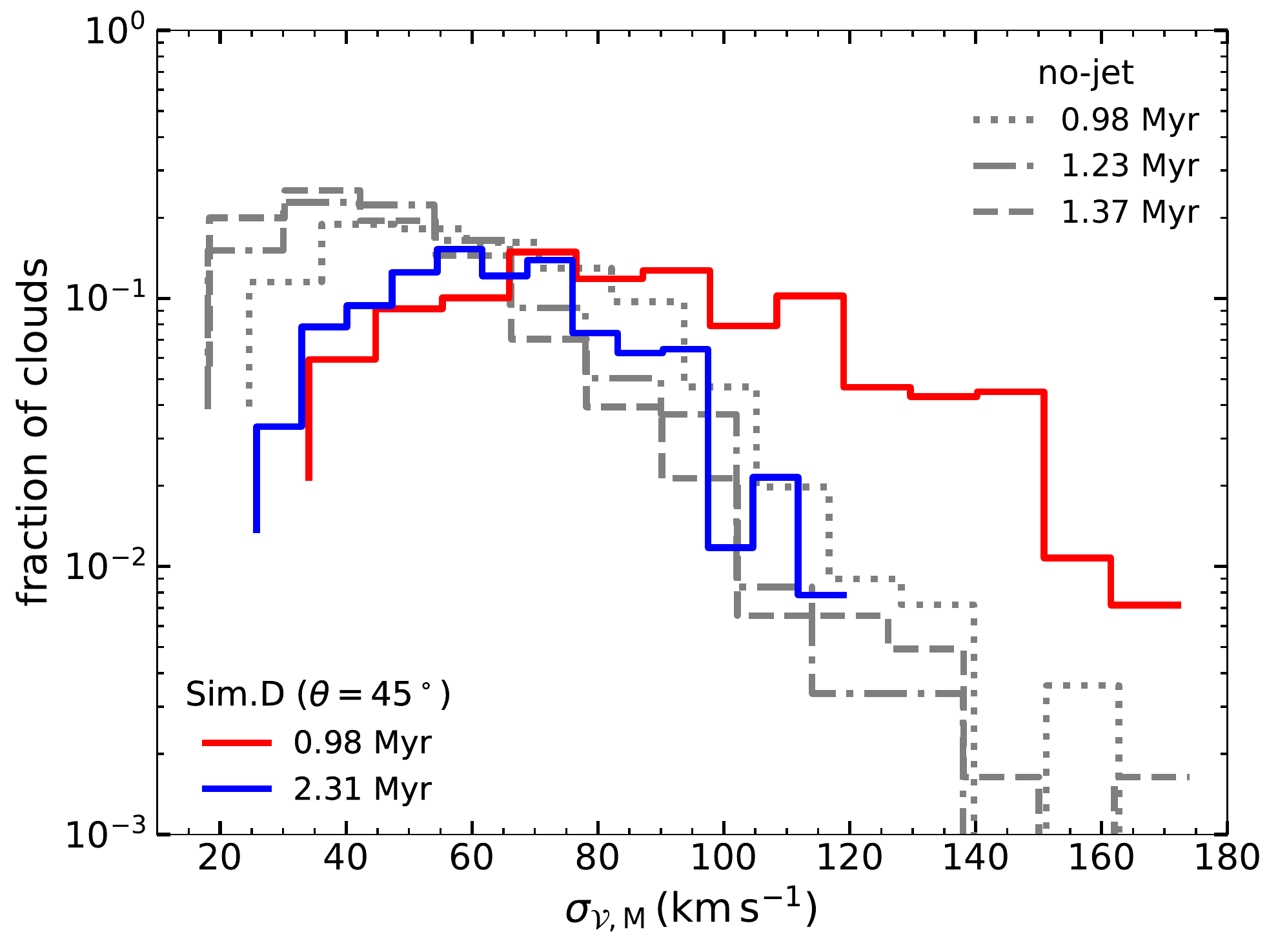} &
  \includegraphics[width=0.5\linewidth]{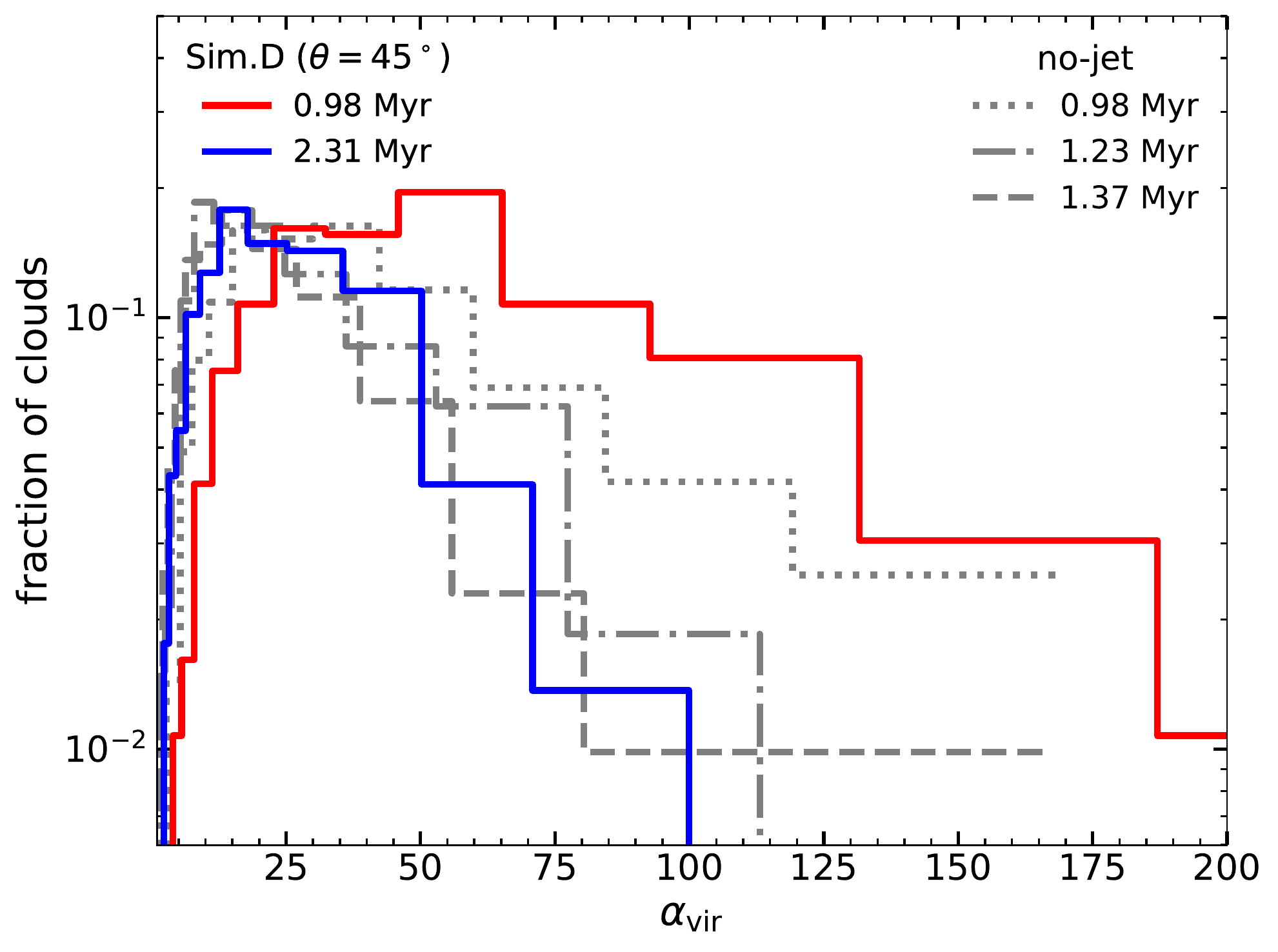}
\end{tabular}}
  \caption{Distribution (number fraction of the clouds) of mass-weighted velocity dispersion (left) and virial parameter (right) for Sim.~D. The different grey lines in both the panels correspond to the respective results for the `no-jet' simulation at different times. The red and blue solid lines show the distribution at 0.98 and 2.31~Myr for Sim.~D.}
  \label{sig_alpha pdf sim D}
\end{figure*}
\subsubsection{Evolution of the density PDF}\label{density PDF Sim.D}
The evolution of the density PDF of the gas in the right panel of Fig.~\ref{density pdf sim D} illustrates how the jet affects the density distribution of the ISM globally. The left and middle panels show the mid-plane number density maps in the $x\mbox{-}z$ plane at the corresponding times (0.98 and 2.31~Myr respectively). The grey lines are density PDFs for the `no-jet' simulation at different times. For the `no-jet' simulation, once the disc settles down, the PDF does not significantly vary with time. For Sim.~D, we see that at 0.98~Myr, the jet is still contained inside the disc. Some of the jet energy leaks through narrow channels, creating a high-pressure bubble. However, at 2.31~Myr, the jet has escaped fully from the disc, sweeping out gas along its path. The density PDF shows an increase in both high- and low-density regions, compared to the `no-jet' simulation. The compression induced by the jet enhances the density in some dense clumps, and the strong shocks remove the material from the ambient medium, creating gas-depleted pockets. Similar trends of the density PDF of an ISM undergoing strong interaction with a jet have been demonstrated previously in \citet{Sutherland_2007,Mukherjee_2016,Mukherjee_2017}.

\subsubsection{Evolution of the turbulent velocity dispersion}\label{turbulence evolution Sim.D}
\begin{figure*}
    \centering
    \includegraphics[width=\linewidth]{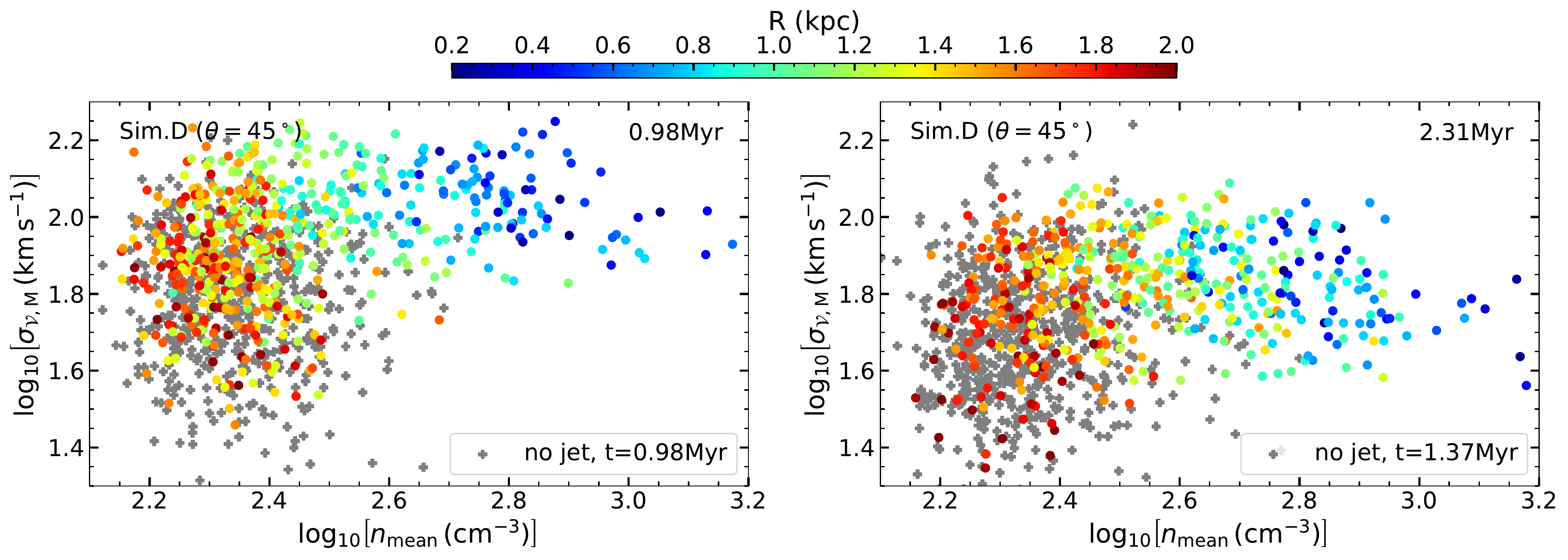}
    \caption{Distribution of mass-weighted velocity dispersion ($\sigmav$) as a function of mean number density ($n_\mathrm{mean}$) of the clouds at 0.98~Myr (left) and 2.3~Myr (right) for Sim.~D. The points are coloured based on the cylindrical distance $R=(x^2_\mathrm{cm}+x^2_\mathrm{cm})^{1/2}$, where $x_\mathrm{cm}$ and $y_\mathrm{cm}$ are the $x$ and $y$ coordinates of the centre of mass vector of a cloud in cylindrical coordinates, measured from the galactic centre. The grey markers are the results of the `no-jet' simulation.}
    \label{n_sigma sim D}
\end{figure*}
Strong shocks from the jet can induce turbulence inside the dense clouds, increasing the velocity dispersion ($\sigmav$) and the virial parameter ($\alphavir$), which is directly related to the turbulent kinetic energy of the gas. In Fig.~\ref{sig_alpha pdf sim D}, we show the distribution (number fraction of the clouds) of $\sigmav$ (left) and $\alphavir$ (right) for Sim.~D. The grey lines are the corresponding distributions for the `no-jet' simulation at different times. We notice that for the `no-jet' simulation, the mean $\sigmav$ of the clouds decreases with time due to the decay of turbulence in the absence of a driving source \citep{MacLow_1998,Padoan_1999}. 

For Sim.~D, the mean velocity dispersion of the clouds increases due to the strong jet-ISM interaction, which transfers jet energy into gas kinetic energy. This can be seen from the increased distribution of $\sigmav \gtrsim 120$~\kms at $\sim 0.98$ Myr, more than three times the mean initial value of $\sim 40$~\kms. However, once the jet decouples from the disc after creating a channel through the gas, it has a reduced effect on the disc gas. Due to much lower resistance, most of the jet energy escapes through the channel. In the absence of strong driving, the velocity dispersion decreases due to the decay of turbulence, as can be seen in the $\sigmav$ distribution at 2.31~Myr. Similar behavior can also be seen in the distribution of $\alphavir$. The initial mean value of $\alphavir \sim 20$ increases to $\sim 100$ at 0.98~Myr, when the jet starts to act on the ISM. After the jet breaks out, $\alphavir$ decreases again until it reaches values similar to the `no-jet' case. Thus, we find that there is a strong increase in $\alphavir$ from its initial value by nearly an order of magnitude under the influence of the jet. The disc then subsequently resettles to values of $\alphavir$ similar to the initial phases after the jet breaks out of the disc. Such qualitative behaviour will be expected for any general scenario of a jet-ISM interaction irrespective of the initial conditions of the ISM.

The jet not only drives turbulence but also compresses the medium, enhancing the mean density of the region. Thus, we expect a correlation between the velocity dispersion ($\sigmav$) and the mean density ($\nmean$) of the clouds, at least inside the region of the gas disc directly interacting with the jet. In Fig.~\ref{n_sigma sim D}, we show the distribution of $\sigmav$ as a function of mean density ($\nmean$) of the clouds for Sim.~D. The colours represent the distances of the clouds from the centre of the galactic disc. The grey markers are the corresponding results for the `no-jet' simulation. We notice that the clouds in the central region ($R<1~\mathrm{kpc}$) are strongly affected by the jet. Such clouds show both an increase in $\sigmav$ and in the mean density due to the compression from the jet. The clouds on the outskirts ($R>1~\mathrm{kpc}$) also show some increase in $\sigmav$, which primarily arises from the shocks percolating through the fractal ISM and backflows from the large-scale bubble. However, the relative increase is milder than for the clouds in the central region directly interacting with the jet. Nonetheless, at later times, as the jet's region of influence increases, the clouds beyond the central 1~kpc also show an increase in density along with a moderate increase in $\sigmav \sim 60~\mathrm{kms^{-1}}$, depicted by the horizontal branch in the right panel of Fig~\ref{n_sigma sim D}.

\subsection{Star formation rate (SFR)}
\subsubsection{Evolution of the global SFR}
\begin{figure}
    \centering
    \includegraphics[width=\linewidth]{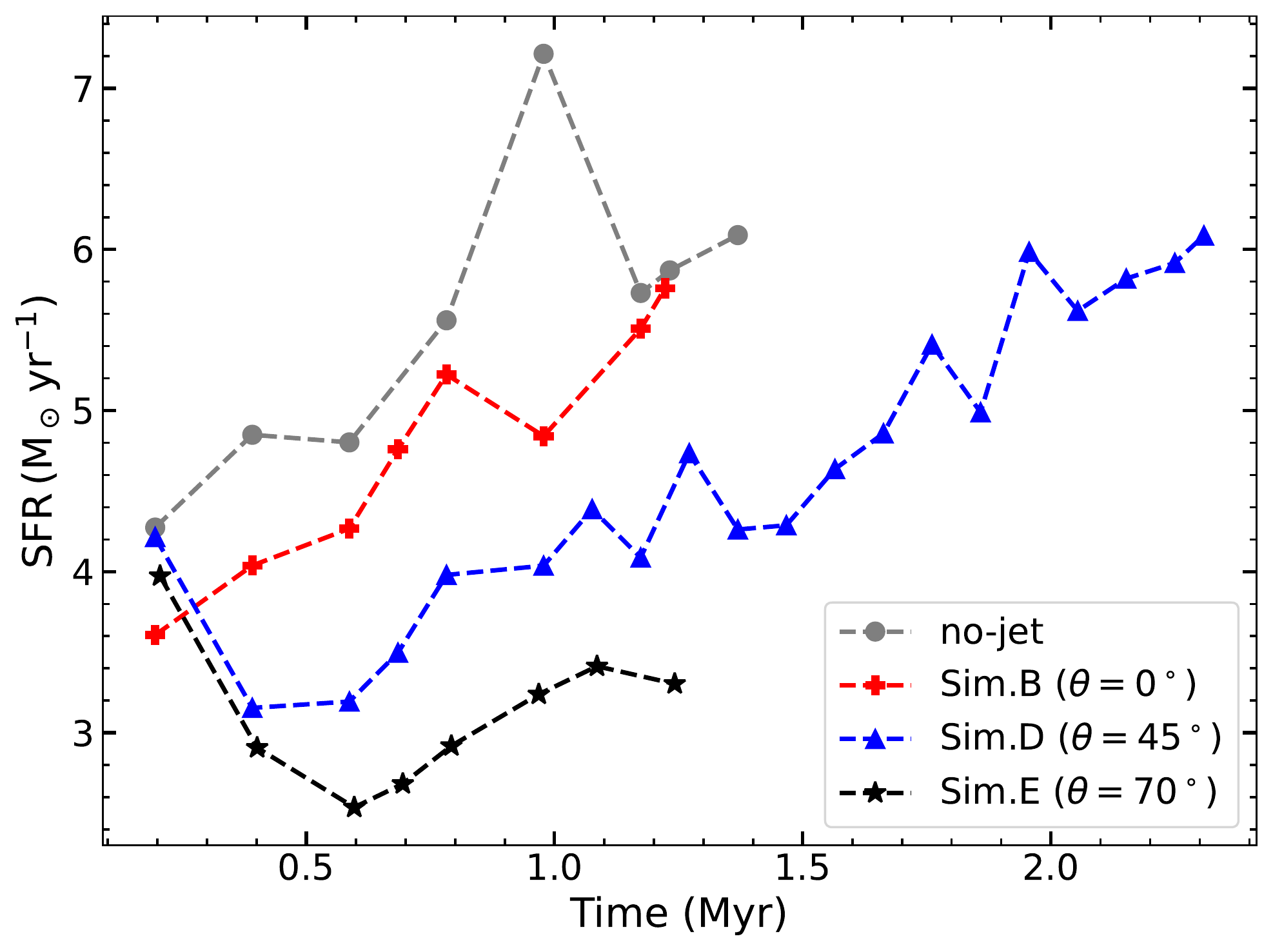} 
    \caption{Evolution of the global star formation rate (SFR) with time for all the simulations. The grey, red, blue and black lines correspond to the `no-jet', B, D and E simulations, respectively.}
    \label{global SFR}
\end{figure}
In this section, we discuss the effect of jet feedback on the evolution of the global SFR. The total SFR depends on how strongly the jet interacts with the gas before breaking out of the disc. In Fig.~\ref{global SFR}, we show the evolution of total (global) SFR with time for all the simulations. The grey, red, blue, and black lines correspond to the `no-jet', B, D, and E simulations.

We notice that all the jetted simulations have a reduction in the global SFR by a factor of a few, depending on the specific simulation (e.g., a factor of $\sim 2$ for Sim.~E) compared to the no-jet simulation, irrespective of the jet inclination angle. This reduction of SFR compared to the no-jet simulation is likely due to the increase in local velocity dispersion and temperature (see Fig~\ref{sig_alpha pdf sim D}) when the jet progresses through the ISM, injecting kinetic and thermal energy into the gas. As a result, the virial parameter of the star-forming clouds increases, which causes a reduction in SFR. Further detailed effects of the jet inclination will be discussed in Sec.~\ref{jet inclination}.
\begin{figure*}
    \centering
    \includegraphics[width=\linewidth]{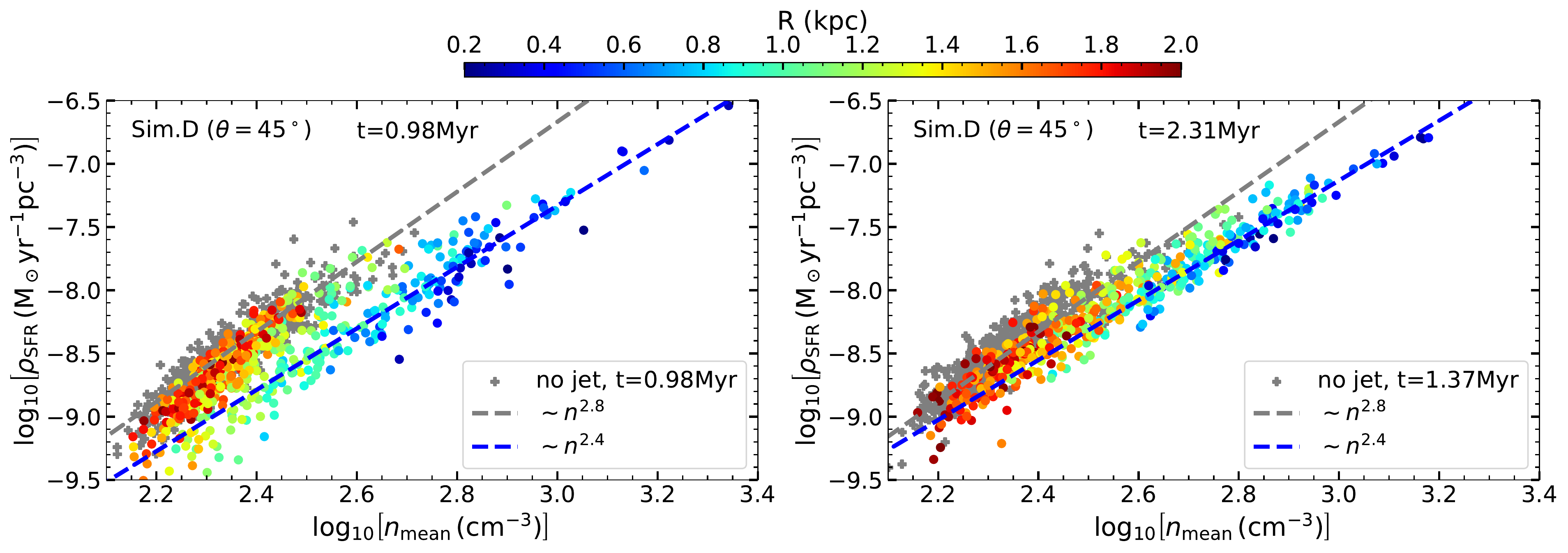}\\
    \includegraphics[width=\linewidth]{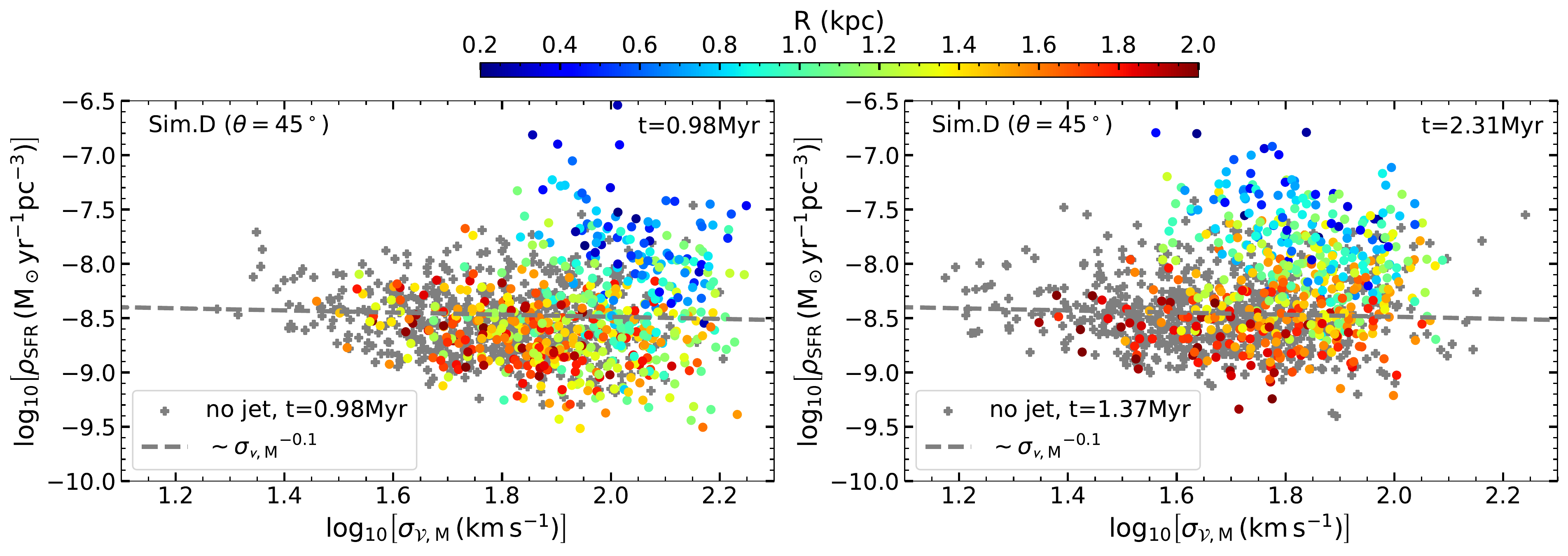}
    \caption{\textit{Top:} SFR density ($\rhoSFR$) as a function of mean number density ($\nmean$) for each cloud at 0.98~Myr (left) and 2.31~Myr (right) for Sim.~D colour-coded by the distances of the clouds from the galactic centre. The grey and blue lines are power-law ($\rhoSFR\sim \nmean^\kappa$) fits to the no-jet points and the clouds in the central region ($R<R_\mathrm{jet}$) for the jetted simulation, where $R_\mathrm{jet}$ is defined as the cylindrical radius beyond which the clouds are not expected to be directly affected by the jet. The value of $R_\mathrm{jet}$ is found to be 0.9 and 1.1~kpc for Sim.~D at 0.98 and 2.31~Myr by visual inspection of the density slice plot in the $X\mbox{-}Z$ plane (Fig.~\ref{density pdf sim D}). The values of $\kappa$ are shown in the legends. \textit{Bottom:} $\rhoSFR$ as a function of mass-weighted velocity dispersion ($\sigmav$) at 0.98~Myr (left) and 2.31~Myr (right) for Sim.~D. The grey dashed line is the fit to a power law ($\rhoSFR\sim\sigmav^{\zeta}$) to the grey points for the no-jet simulation. The value of $\zeta$ is indicated in the legend.}
    \label{n_SFR sim D}
\end{figure*}
We also notice that, for a particular simulation, the global SFR increases with time after the initial decline (Fig.~\ref{global SFR}). This is due to density enhancements from shocks, turbulence decay, and cooling of the dense clouds. However, the rate of change differs for different simulations, which depends on how strongly the jet interacts with the gas, the duration for which the interaction is sustained, and to what spatial extent the jet affects the gas. As discussed before, once the jet breaks out of the disc, it becomes much more inefficient in driving turbulence and heating the gas inside the disc. However, the density enhancement due to compression creates dense star-forming cores. When the jet decouples from the disc, the dying turbulence and increased density make the clouds more efficient in star formation. This causes the increase in SFR after the initial decline, as seen in Fig.~\ref{global SFR}.

\subsubsection{Trends in SFR for individual clouds}\label{individual SFR}
\begin{enumerate}
\item[(a)] \textbf{Volumetric SFR density ($\rhoSFR$):} \\
In this section, we discuss how the SFR inside individual clouds is affected by the jet. The efficiency of star formation inside the dense clouds is regulated by their interaction with the jet. Here, we divide the clouds into two categories depending on how they are affected by the jet: (i) clouds inside a cylindrical radius $R_\mathrm{jet}$ are directly interacting with the jet, (ii) clouds outside $R_\mathrm{jet}$, which are not directly interacting with the jet. Nevertheless, the clouds in the outskirts can be affected by the backflows from the jet, which can potentially affect the large-scale disc. For each simulation snapshot, we evaluate $R_\mathrm{jet}$ by visually inspecting the spatial extent of the jet in the $X\mbox{-}Z$ plane.

The top panels of Fig.~\ref{n_SFR sim D} shows the volumetric SFR density ($\rhoSFR$) as a function of $\nmean$ at 0.98~Myr (left) and 2.31~Myr (right) for Sim.~D. The grey circles are the corresponding $\rhoSFR$ for the `no-jet' simulation. From Fig.~\ref{density pdf sim D}, we find that $R_\mathrm{jet}=$ 0.9 and 1.1~kpc at 0.98 and 2.31~Myr for Sim.~D. The grey and blue lines are fits to the no-jet data points and the central clouds ($R<R_\mathrm{jet}$) for Sim.~D by a power law ($\rhoSFR\sim \nmean^\kappa$), respectively. 

We notice a bi-modal distribution of $\rhoSFR$ at 0.98~Myr when the jet is confined in the disc, which represents the two roles that the jet-induced turbulence has for the gas. First, it leads to a strong compression of clouds that are near the jet axis, as indicated by the generally higher gas densities of blue dots in Fig.~\ref{n_SFR sim D} than in the no-jet simulation (grey dots) and gas further away from the jet in Sim~D (red dots). Individual clouds that are very near the jet axis also reach high SFRs that are above those reached by any of the clouds further out in the jet simulation, and in the comparison simulation without jet.
\begin{figure*}
\centerline{
\def\arraystretch{1.0}
\setlength{\tabcolsep}{0.0pt}
\begin{tabular}{lcr}
     \includegraphics[width=0.33\linewidth]{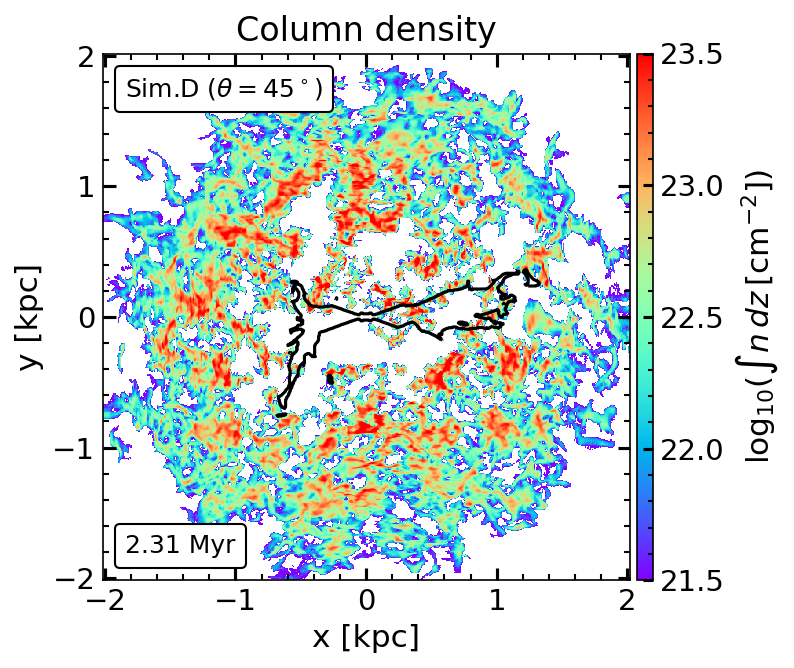}
     & \includegraphics[width=0.33\linewidth]{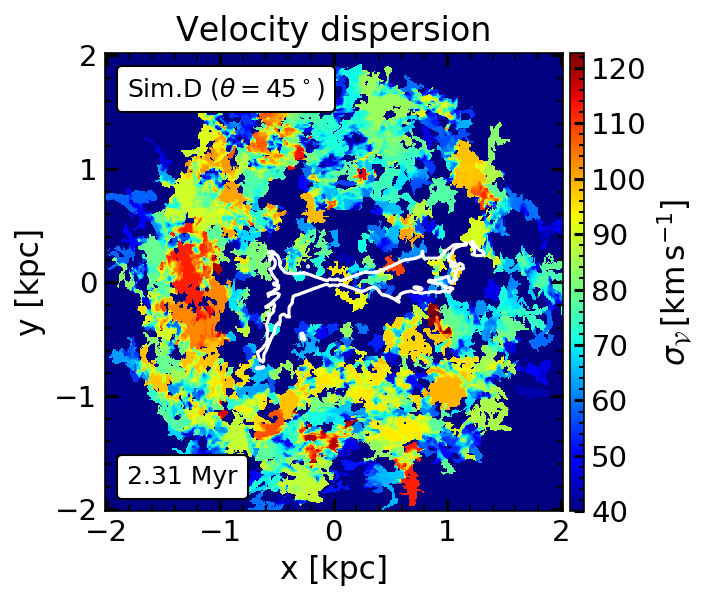}
     & \includegraphics[width=0.33\linewidth]{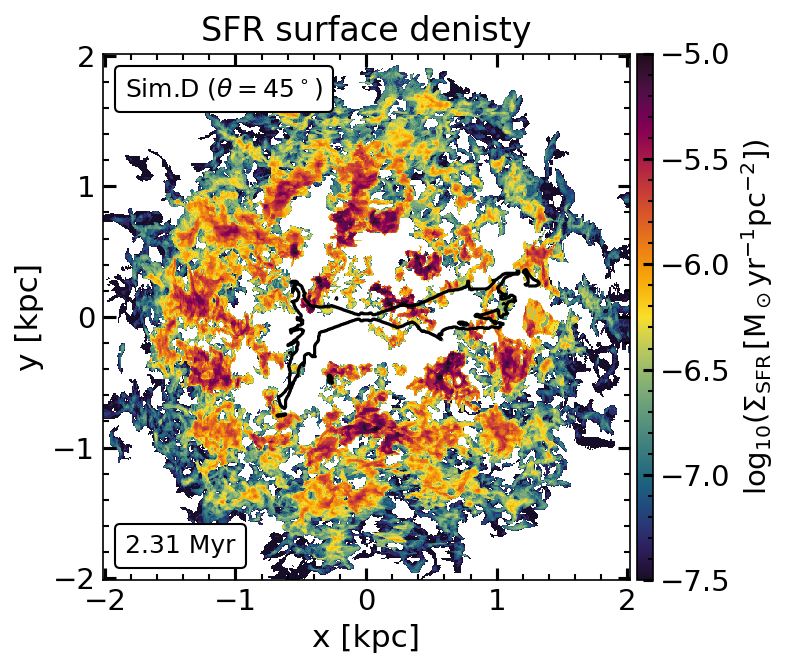}
\end{tabular}
}
\caption{\textit{Left:} The column density map for Sim.~D at 2.31~Myr including the mass inside the clouds only. \textit{Middle:} The average 3D velocity dispersion of the dense clouds in the $x\mbox{-}y$ plane. \textit{Right:} The SFR surface density in the $x\mbox{-}y$ plane in units of $\mathrm{M_\odot\,yr^{-1}\,pc^{-2}}$. The black lines in the left and right panel and the white line in the middle panel are the contours of the jet tracer at 0.5 (where a value of 0 and 1 would correspond to no-jet and jet-only material, respectively) projected onto the $x\mbox{-}y$ plane by a volume-weighted average along the $z$-axis. For evaluating the maps of velocity dispersion and SFR surface density, the corresponding value for each cloud is first found, and the same is assigned to all pixels inside a cloud. The distribution of $\sigmav$ in the $x\mbox{-}y$ plane is then obtained by evaluating the mass-weighted mean along the $z$-axis. The SFR distribution is found by adding the SFR values for each pixel along the $z$-axis.}
\label{SFR map}
\end{figure*}
However, globally, this blue sequence falls below that set by the control simulation and outskirts in the jet simulation (see also Fig.~\ref{global SFR}). This shows that at a given gas density, the efficiency of turning gas into stars is lower in the clouds that are strongly affected by the radio jet than in clouds that reach similar densities without jet compression. This can be explained by the second role of turbulence for the dense gas, which not only compresses the gas but also enhances the turbulent velocity dispersion within the clouds. We see that the blue line (fit to the clouds inside $R<0.9\,\mathrm{kpc}$) is almost an order of magnitude lower than the grey line. Thus, this offset is indicative of the fact that turbulence reduces the efficiency of star formation in the clouds that show positive feedback. However, we must note that the turbulence is induced throughout the disc, leading to a global reduction in SFR. Thus the global impact of the radio jet is to lower the overall SFR within the galaxy whereas simultaneously enhancing the SFR in local clouds close to the jet.

Our results are supported by a number of observational studies, which have shown in the past that a significant subset of radio galaxies host large amounts of moderately dense molecular and highly turbulent gas stirred up by the radio jets \citep[][]{Ogle_2007,Ogle_2010,Alatalo_2011,Nesvadba_2010,Nesvadba_2011,Lanz_2015}, which however do not induce star formation at the rates typically observed at the same gas-mass surface densities as in galaxies without powerful jets. These studies have also found characteristic offsets in the KS diagram similar to what we find here \citep[][]{Nesvadba_2010,Ogle_2010,Nesvadba_2011}. \citet{Nesvadba_2011} showed that the observed line broadening on kpc scales is consistent with the observed low SFRs, if turbulence not only causes the observed high gas velocity dispersion on kpc scales, but also creates a turbulent cascade that dominates the gas kinematics on the scales of individual molecular clouds. The presence of jet-induced, low-efficiency star formation has also been shown in observations of gas clouds in typical examples of jet-induced star formation like in Centaurus A \citep{Salome_2017} and potentially Minkowski's Object \citep[][]{Lacy_2017}. It may also be the possible origin of systems where star formation is found to be aligned with the radio jet \citep[e.g.,][]{Dey_1997,Bicknell_2000,Klamer_2004,Privon_2008}. Positive feedback in such galaxies does not seem to enhance the global star formation above what is observed in equally gas-rich, massive, dusty star-forming galaxies at similar redshifts \citep[][]{Man_2019,Nesvadba_2020}.

However, at 2.31~Myr, when the jet decouples from the disc, this bi-modality goes away. This is due to the absence of strong interaction and the resultant decay of the velocity dispersion, which makes the density-enhanced clouds relatively more efficient in forming stars. This can be seen from the upper right panel of Fig.~\ref{n_SFR sim D}, where we see that at 2.31~Myr, the clouds inside the central region have moved towards the grey line, reducing the offset compared to the earlier time. There is still a hint of reduced efficiency when compared to the undisturbed clouds, as seen from the lower value of the power-law index of the fit. Thus, we see that the timescale of direct interaction of the jet is a major factor in regulating the SFR.

We show the two-dimensional distribution of the column density, 2D velocity dispersion ($\sigmav$) and SFR surface density in Fig.~\ref{SFR map} to highlight the morphology of the star-forming sites. We see that the jet has created a central cavity by shredding the gas in the central region ($\sim 1$ kpc). However, there is also an increased column density and velocity dispersion in patches around the central cavity. Interestingly, the velocity dispersion increases considerably near the immediate vicinity of the jet head, at the strongest point of interaction between the jet stream and the ISM. In the SFR surface density map, we see enhanced star formation at these locations. The gas near the central region experiences the most compression from the jet as this is the region with the strongest interaction. As a result, we see a ring-like area of enhanced SFR, which has also been proposed in other theoretical studies \citep{gaibler_2012,Dugan_2014,Mukherjee_2018}. A similar kind of ring-like enhanced star-forming region has also been found observationally in galaxies containing an AGN at the centre \citep{Shin_2019}.

\begin{figure}
\centerline{
\def\arraystretch{1.0}
\setlength{\tabcolsep}{0.0pt}
\begin{tabular}{lcr}
  \includegraphics[width=\linewidth]{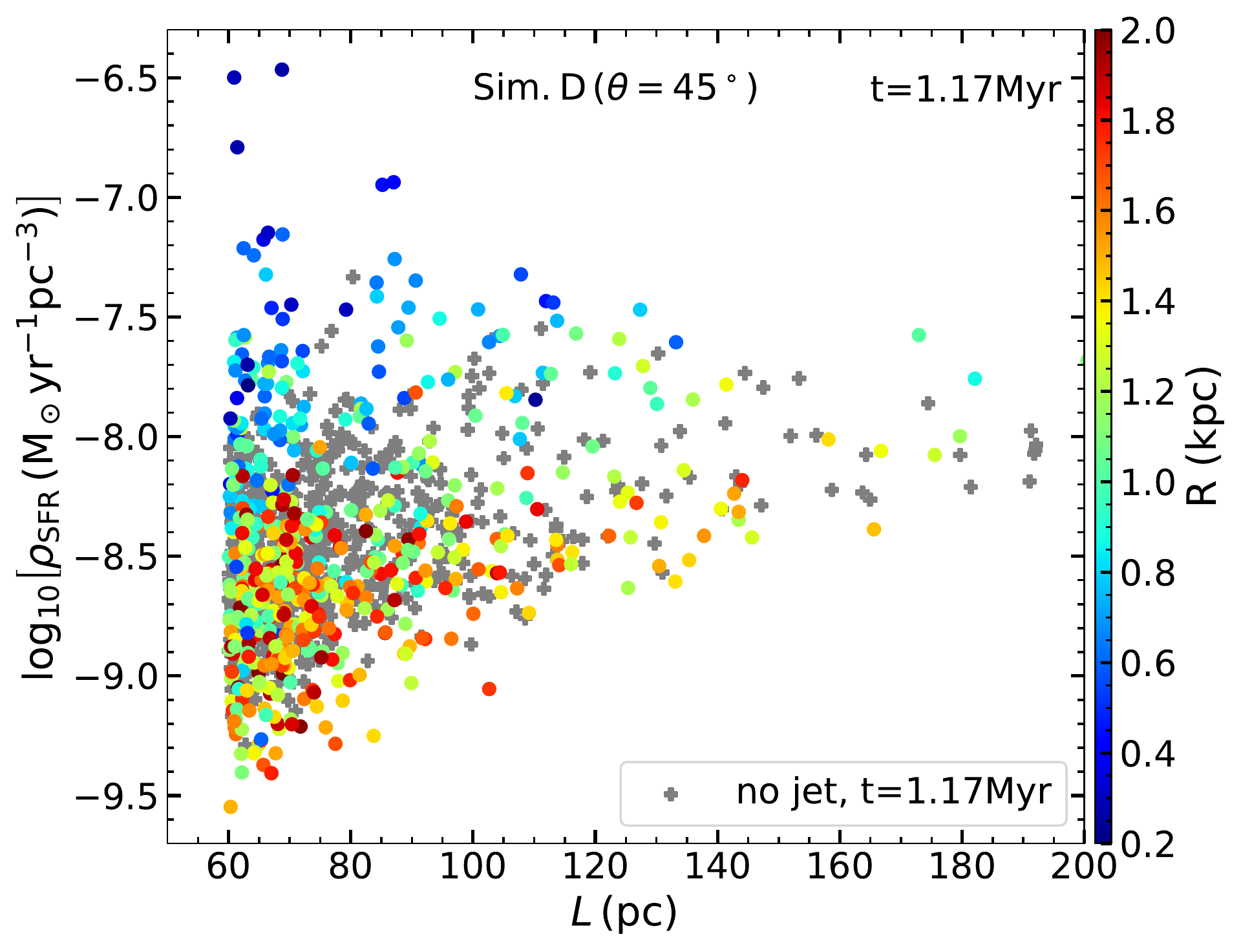} \\
  \includegraphics[width=\linewidth]{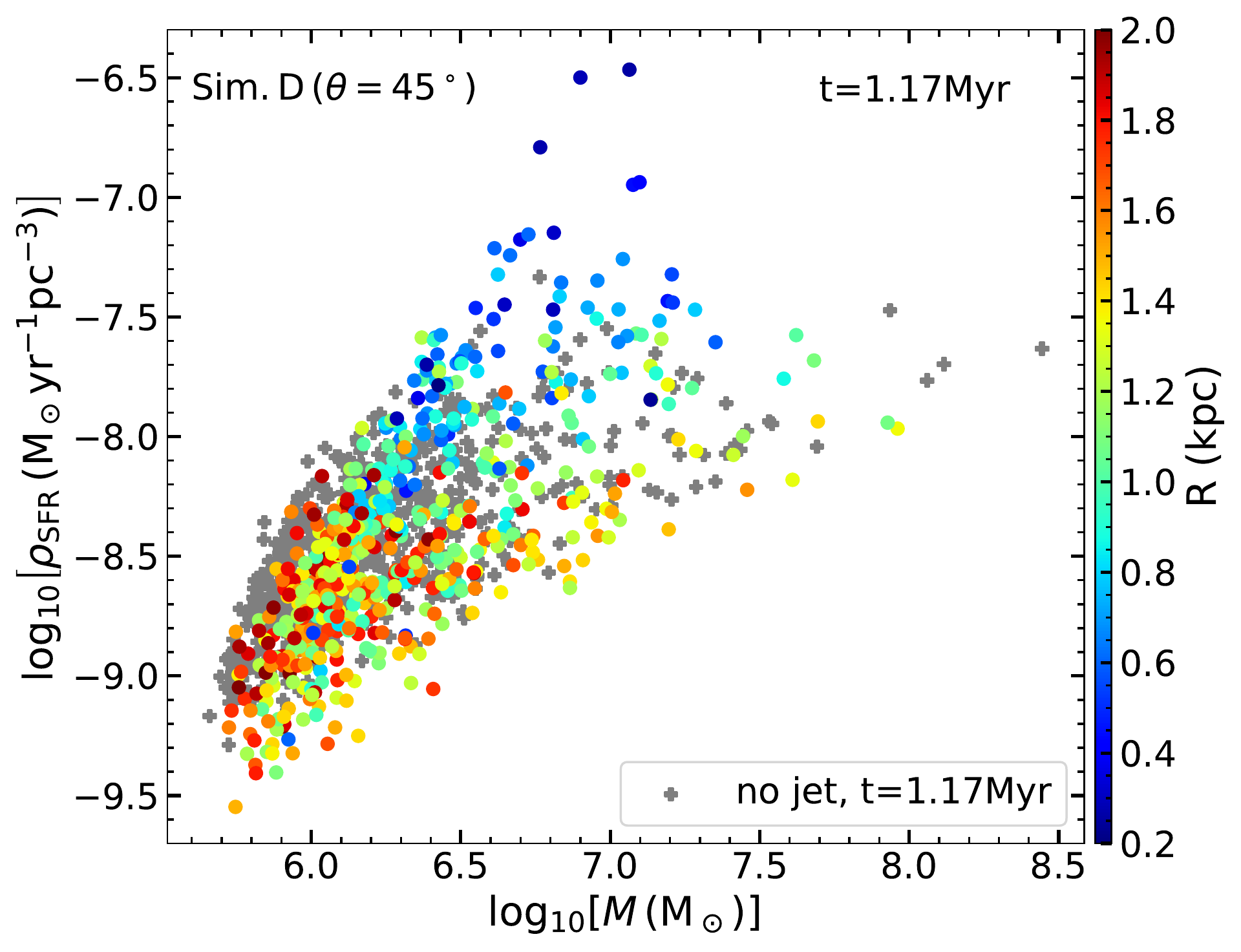}
\end{tabular}}
  \caption{$\rhoSFR$ as a function of the cloud size (upper) and the cloud mass (lower) for individual clouds colour-coded by the distance of the clouds from the centre.}
  \label{size_mass_SFR Sim.D}
\end{figure}

The interpretation of the local positive feedback may seem contradictory when compared to the effect of the jet on the global SFR, where we see an overall reduction of the SFR for the jetted simulations, implying global negative feedback (c.f., Fig.~\ref{global SFR}). However, it is important to note that here we are discussing the volumetric SFR density ($\rhoSFR$), not the SFR itself. The contribution from the individual clouds to the total SFR depends on $\rhoSFR$ as well as the size and mass of the clouds. Most of the clouds that show an enhanced SFR density in the inner region ($R< R_\mathrm{jet}$) also have a smaller size, as the direct impact of the jet shreds the outer layers. This is demonstrated in Fig.~\ref{size_mass_SFR Sim.D}, where we show $\rhoSFR$ as a function of the approximate size of the clouds ($L\approx V^{1/3}$, where $V$ is the volume of a cloud) and the mass of the clouds. The clouds with high SFR density tend to be smaller in size and have intermediate gas mass. Also, the number of such clouds forms a smaller fraction compared to the total cloud distribution in the disc. Thus, such inner clouds showing positive feedback contribute less to the total budget of the SFR of the whole disc. As a result, we get a global reduction in SFR, whereas the feedback from the jet significantly regulates the SFR locally in different regions of the disc, depending on their distances from the jet.

\begin{figure}
    \centering
    \includegraphics[width=\linewidth]{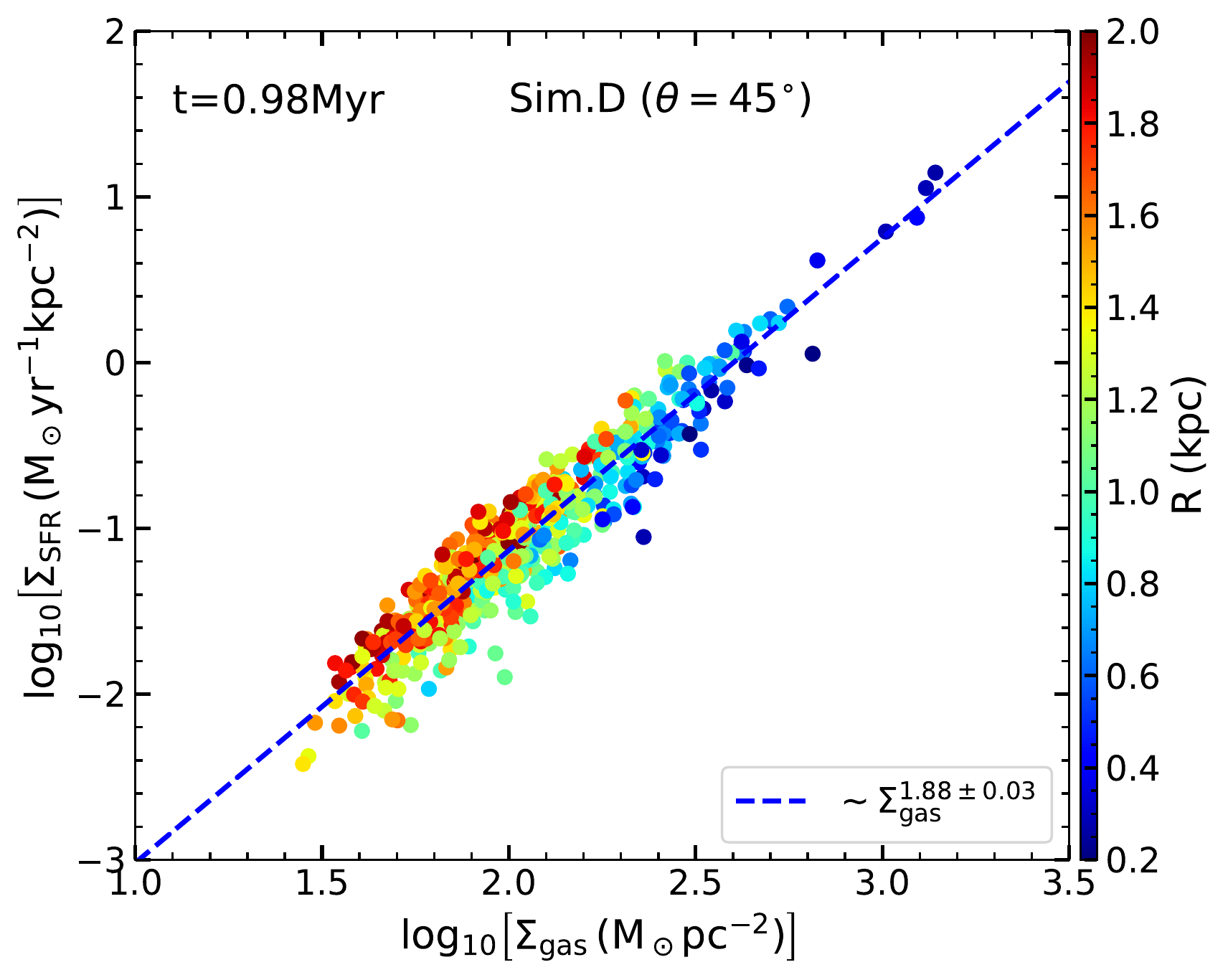}
    \caption{$\Sigma_\mathrm{SFR}$ vs.~$\Sigma_\mathrm{gas}$ for individual clouds for Sim.~D at 0.98~Myr. The black dashed line represent the standard KS relation. The blue dashed line is a power-law fit to the data.}
    \label{SFR_surface Sim.D}
\end{figure}

\item[(b)] \textbf{SFR surface density ($\Sigma_\mathrm{SFR}$):} \\
To further study the properties of the clouds, we have calculated the SFR surface density ($\Sigma_\mathrm{SFR}$) as a function of gas surface density ($\Sigma_\mathrm{gas}$) similar to what we have done in Fig.~\ref{Stellar surface density no-jet}. The results are presented in Fig.~\ref{SFR_surface Sim.D}, where we find again that the values are highly correlated. We also notice that the clouds in the central region ($<1~\mathrm{kpc}$) have higher values of $\Sigma_\mathrm{gas}$ as well as $\Sigma_\mathrm{SFR}$ due to the compression. The slope of $\sim 1.88$ also agrees well with recent high-resolution extragalactic observations \citep[e.g.][]{Liu_2011,Gao_2021}, who find a similar value of $1.96$.
\end{enumerate}

\begin{figure}
  \centering
  \includegraphics[width=\linewidth]{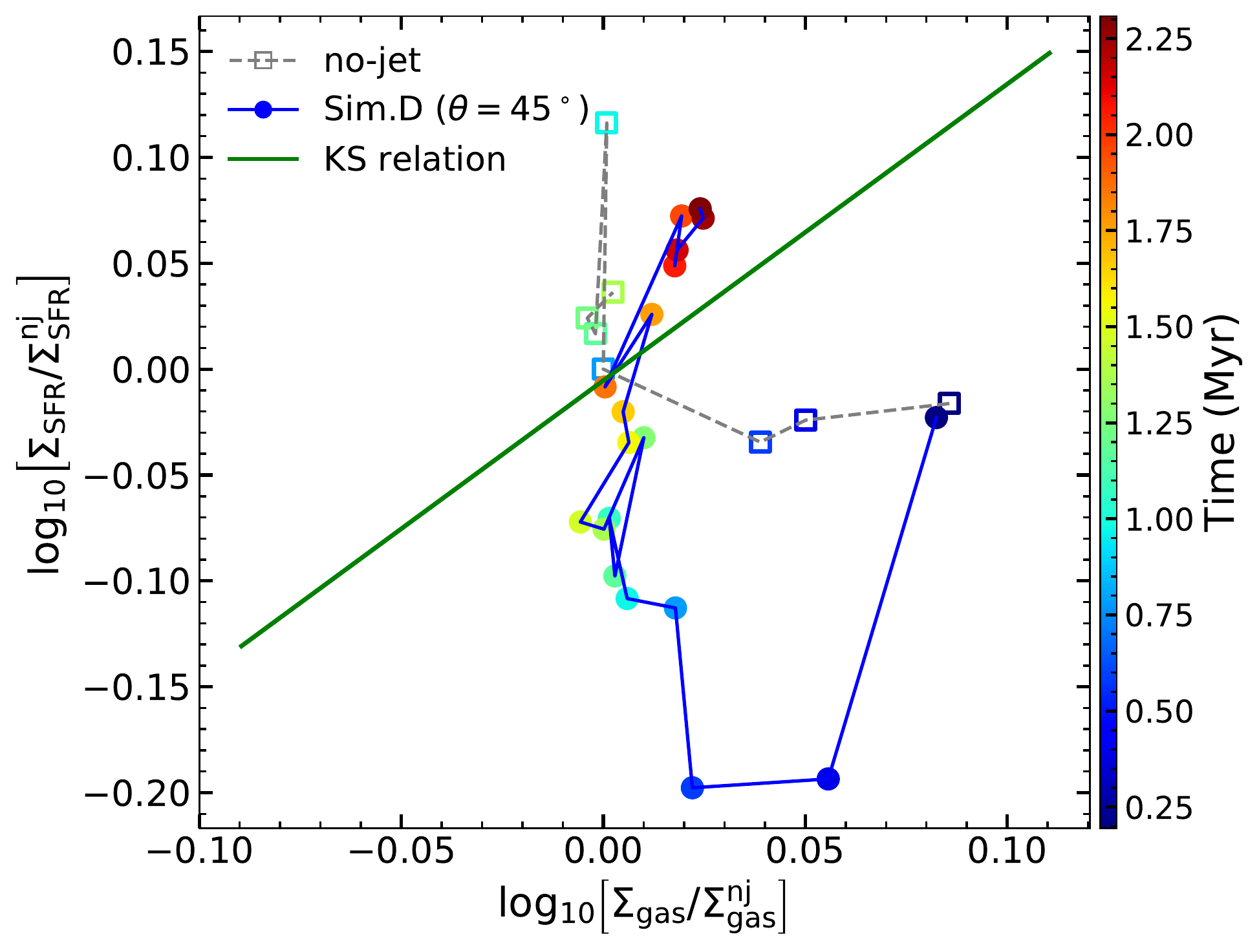} 
  \caption{The evolution of the galaxy in Sim.~D on the normalized Kennicutt-Schmidt diagram. The SFR surface densities and the gas mass surface densities have been normalized with respect to the corresponding values of the `no-jet' simulation at 0.78~Myr. The open squares are the position of the galaxy in the `no-jet' simulation at different times indicated in the colourbar. The coloured filled circles correspond to the gas disc in Sim.~D at different times. The solid green line represents the KS relation normalized by the same normalization constant.}
  \label{KS sim D}
\end{figure}
\begin{figure*}
\centerline{
\def\arraystretch{1.0}
\setlength{\tabcolsep}{0.0pt}
\begin{tabular}{lcr}
     \includegraphics[width=0.5\linewidth]{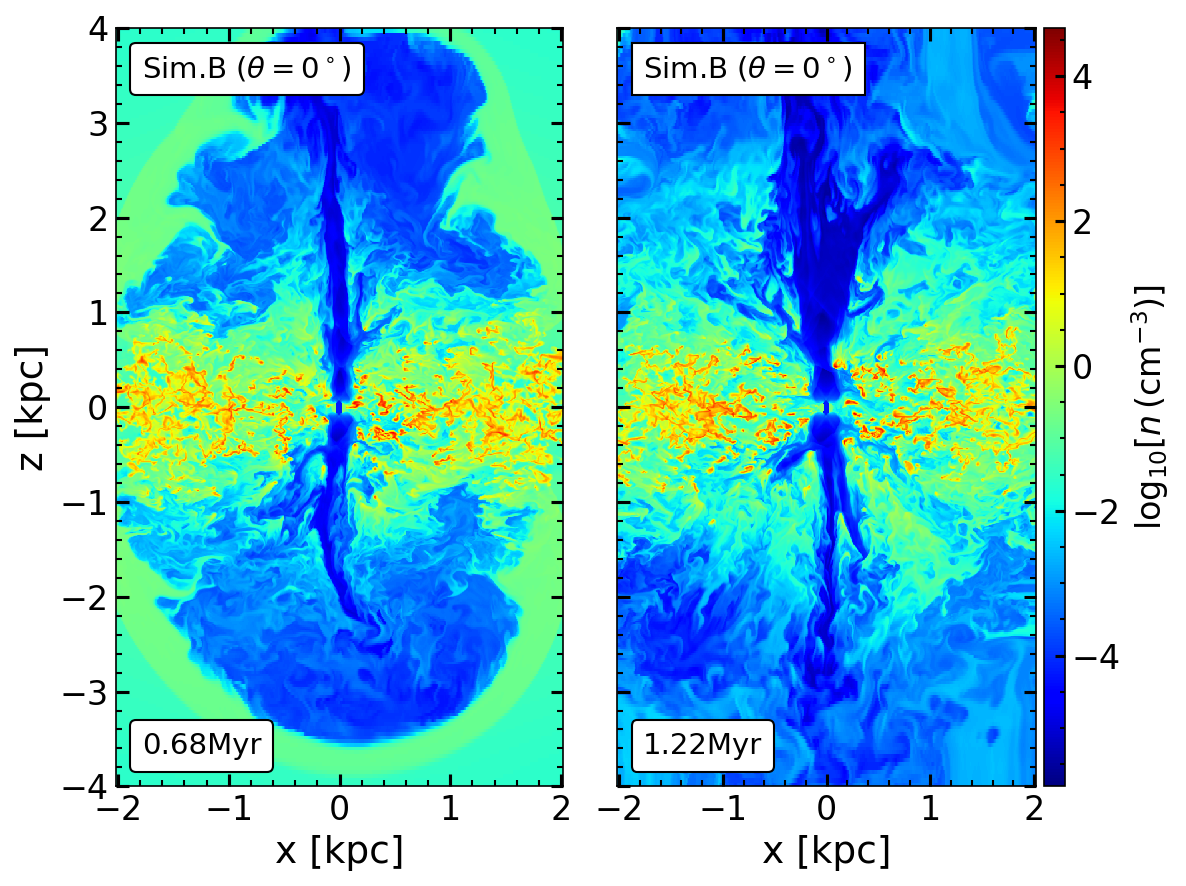}
     & \includegraphics[width=0.5\linewidth]{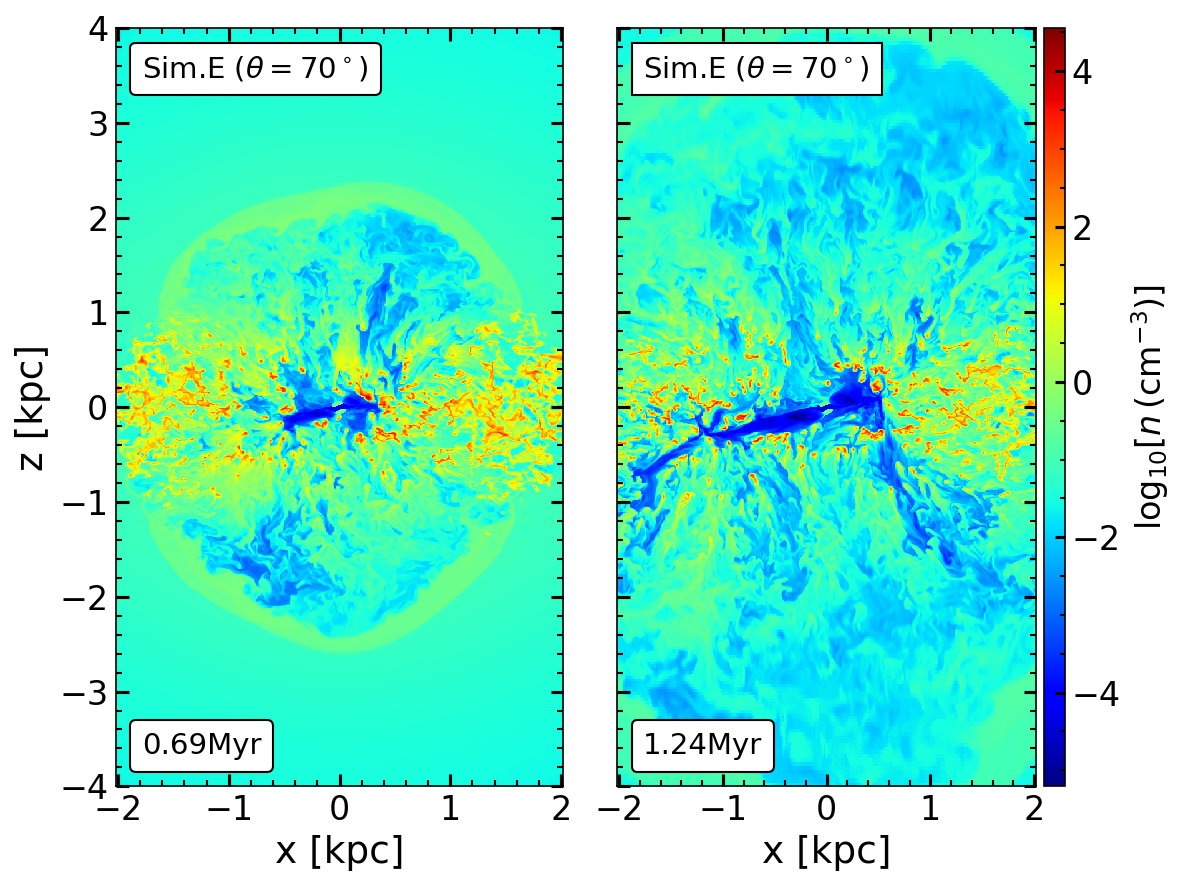}
\end{tabular}
}
\caption{Slice through the number density distribution in the $x\mbox{-}z$ plane for Sim.~B (first and second panels) and E (third and fourth panels) at different times (see labels).}
\label{slice BE}
\end{figure*}
\subsubsection{Time evolution of SFR surface density}
Observationally, the star formation activity in a galaxy is often characterised by comparisons with the KS relation \citep{KS_1998a,KS_1998b}. Here, in Fig.~\ref{KS sim D}, we show the evolution of SFR surface density ($\Sigma_\mathrm{SFR}$) as a function of gas mass surface density ($\Sigma_\mathrm{gas}$) with time for Sim.~D. The surface densities are normalized with respect to the corresponding values of the `no-jet' simulation at 0.78~Myr to quantify the relative evolution of the SFR in the KS diagram. The filled circles correspond to Sim.~D, and the open squares represent the `no-jet' simulation. The data points are coloured by the runtime of the simulation. The solid green line is the normalized KS relation. 

We see that initially, both the `no-jet' and Sim.~D start from the same location in the KS plot. The SFR of the `no-jet' simulation then increases with time. We notice that the $\Sigma_\mathrm{gas}$ evolves slightly in the KS diagram as the gas settles in the disc and the clouds get dispersed. Note that the `no-jet' simulation at $\sim 0.8$~Myr lies exactly on the KS line, as per design, due to our method of calibrating the SFR efficiency ($\epsilon$) in Eq.~\eqref{eq:SFRff}. However, for Sim.~D, the SFR surface density initially decreases as the jet starts to interact with the ISM. After the initial drop, the SFR then increases with time as the density of the clumps increases due to gas compression. This is exacerbated by the decline in velocity dispersion after the jet breaks out of the disc, as discussed earlier in Sec.~\ref{turbulence evolution Sim.D}. The SFR density increases further, and the gas disc starts to move towards the KS line, eventually crossing it. Thus, the expected increase in SFR due to the density enhancement is balanced by the simultaneous increase in velocity dispersion and therefore an increase in $\alphavir$ which keeps the gas disc in simulation D at efficiencies close to the standard KS relation. We note here that the magnitude of variations of the SFR and gas surface densities are small for simulation D (within 1~dex), and so too for the other simulations, as discussed later. However, the qualitative trend of the time evolution in the KS plot shows that the same gas disc can potentially moves through different evolutionary phases, depending on the intensity of the jet-ISM interaction.

\subsection{Effect of Jet Inclination}\label{jet inclination}
The angle of inclination ($\theta$) of the jet launch axis with respect to the disc determines how severely the jet affects the ISM before breaking out of the disc and for how long. In this section, we present results for Sim.~B ($\theta=0^\circ$) and Sim.~E ($\theta=70^\circ$) to see how the jet inclination regulates the behaviour of different physical quantities. We discuss the main results below.

\begin{enumerate}
\item The vertical jet in Sim.~B quickly breaks out of the disc, creating a cone-like structure, removing the gas along its path. At later stages of the evolution, the disc almost remains unaffected by the jet (first and second panels of Fig.~\ref{slice BE}). However, in Sim.~E, the jet encounters a large column density along its path and strongly interacts with the ISM. As the jet proceeds through the disc, it decelerates and becomes sub-relativistic \citep[as also shown earlier in][]{Mukherjee_2018}, launching outflows through small channels (fourth panel of Fig.~\ref{slice BE}).

\begin{figure*}
\centerline{
\def\arraystretch{1.0}
\setlength{\tabcolsep}{0.0pt}
\begin{tabular}{lcr}
  \includegraphics[width=0.5\linewidth]{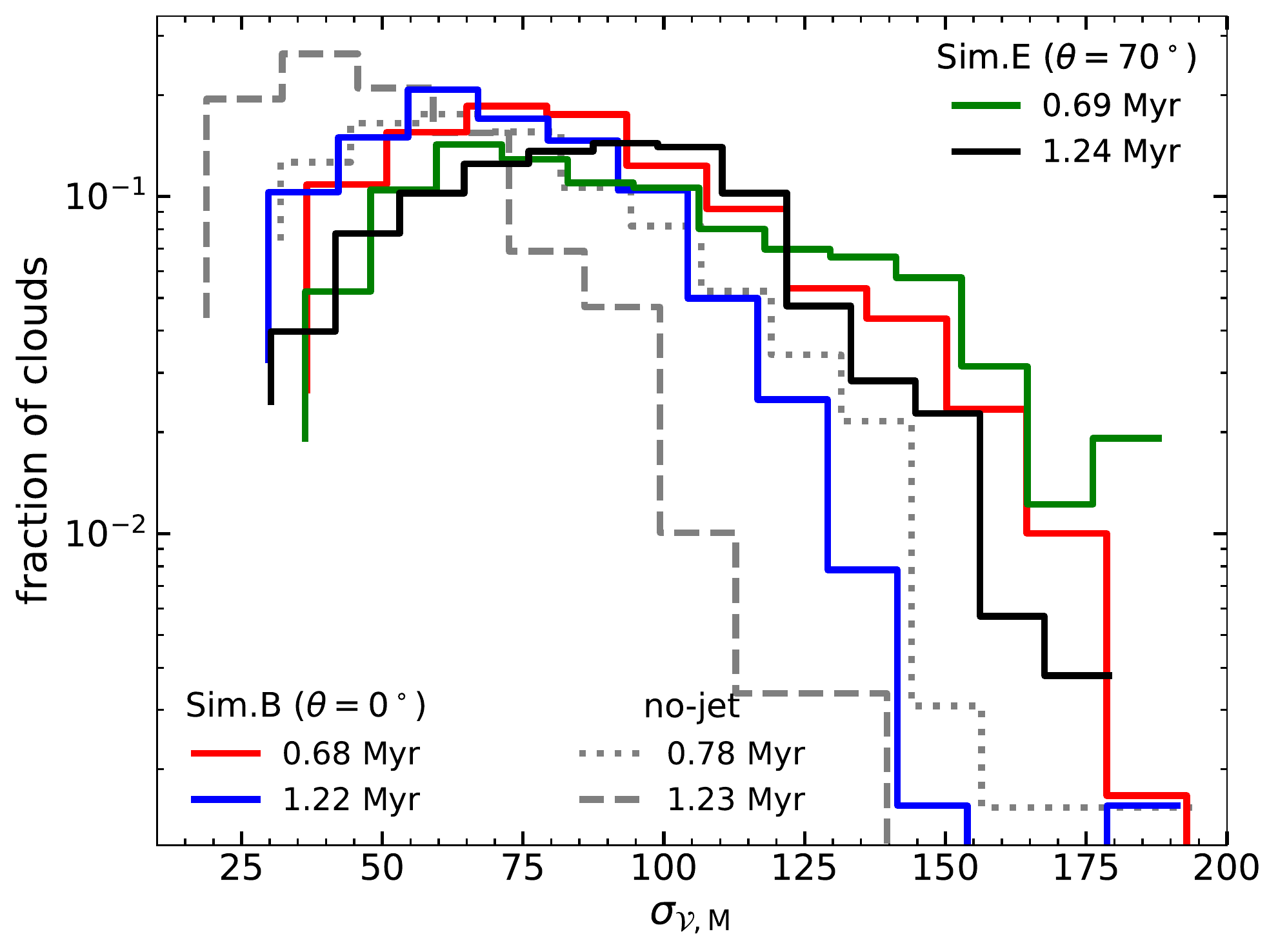} &
  \includegraphics[width=0.5\linewidth]{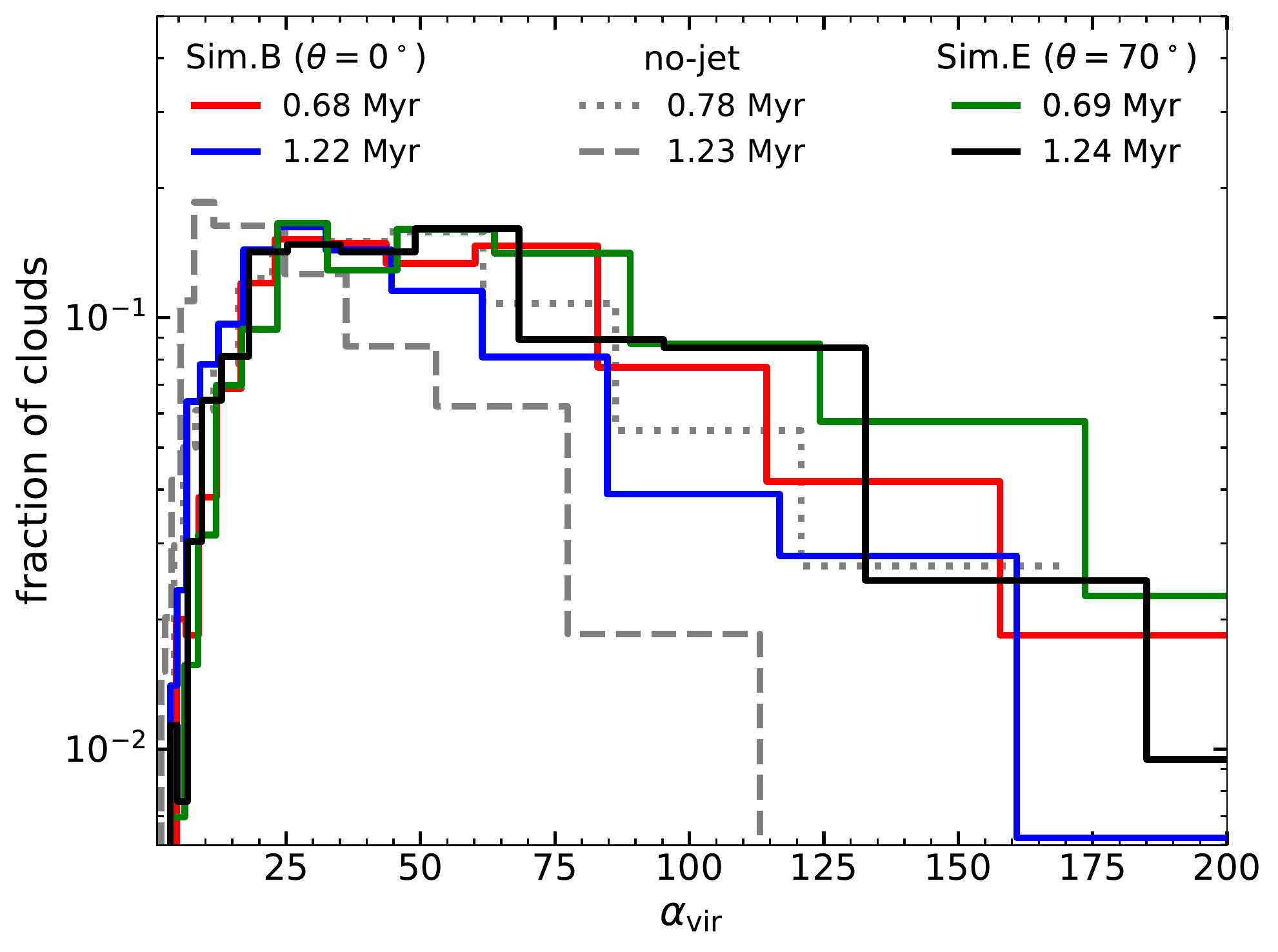}
\end{tabular}}
  \caption{Distribution of mass-weighted velocity dispersion (left) and virial parameter (right) for Sim.~B and E. The grey lines are the corresponding distribution for the `no-jet' simulation at different times. In both panels, the red and green solid lines correspond to $\sim 0.68$~Myr for Sim.~B and E. The blue and black solid lines represent the same at $\sim 1.23$~Myr. We see that for a particular simulation, the mean value of $\sigmav$ decreases from the value at an earlier time. This evolution is similar for $\alphavir$. }
  \label{sig alpha pdf sim BE}
\end{figure*}

\item The distribution of $\sigmav$ (left) and $\alphavir$ (right) for Sim.~B (red and blue) and Sim.~E (green and black) at $\sim 0.69$~Myr and $\sim 1.23$~Myr is shown in Fig.~\ref{sig alpha pdf sim BE}. The grey lines are the corresponding distributions for the `no-jet' simulation at 0.78~Myr and 1.23~Myr. Again, we see that the mean value of $\sigmav$ initially increases, followed by a decline at late times due to decay of turbulence for a given simulation. In general, the clouds in Sim.~E experience a higher velocity dispersion than the clouds in Sim.~B. This is expected, as the jet in Sim.~E strongly interacts with the disc in a larger volume, injecting kinetic and thermal energy into the gas. We also notice that the change in mean velocity dispersion from earlier to later times is less in Sim.~E than in Sim.~B. This is a consequence of the jet in Sim.~E remaining confined inside the disc for a longer time, which replenishes some of the lost energy due to turbulence decay. In Sim.~B, however, the jet quickly decouples from the disc when a channel is created through the gas. Thus, most of the jet energy escapes through the channel without affecting the disc significantly. In the distribution of $\alphavir$ (right panel), we see that the mean value of $\alphavir$ decreases slightly with time for simulation B.

\begin{figure*}
    \centering
    \includegraphics[width=\linewidth]{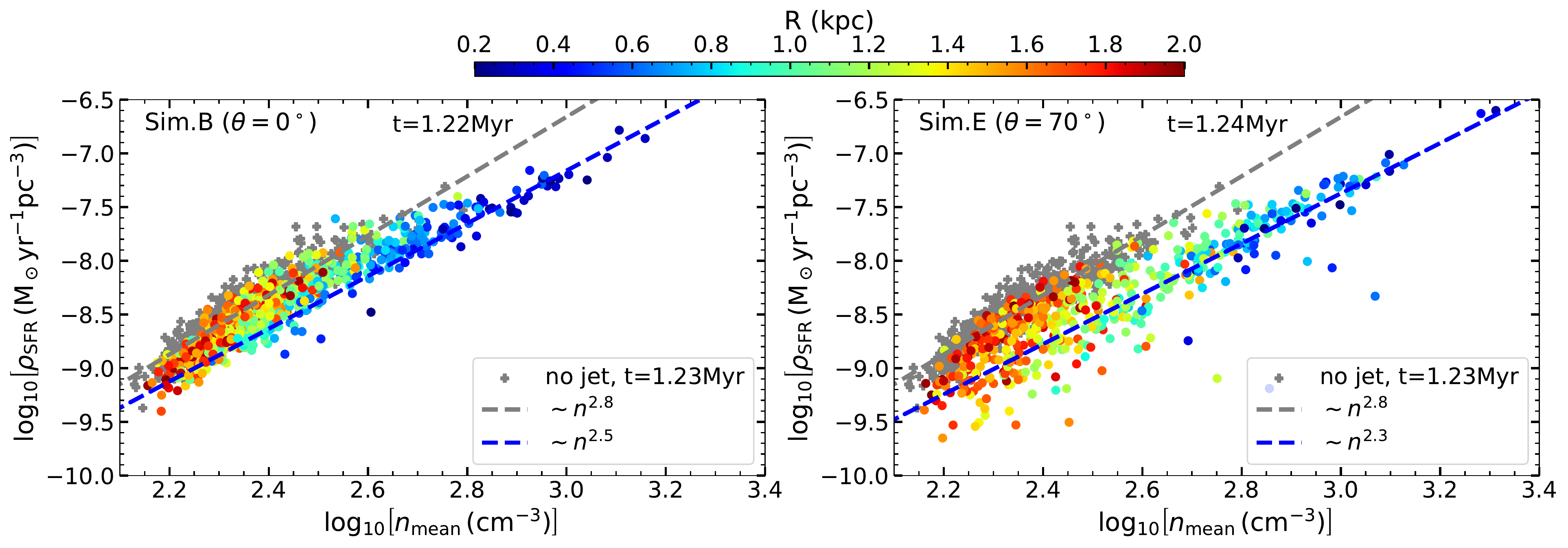}
    \caption{SFR density as a function of the mean number density of individual clouds for Sim.~B (left) and D (right) at time $\sim 1.23$~Myr. The points have been colour-coded with their distance from the galactic centre. The grey markers in both the panels are the corresponding no-jet data points. The grey and the red lines are the same as in Fig.~\ref{n_SFR sim D}. Here the $R_\mathrm{jet}$ values for Sim.~B and Sim.~E at $\sim 1.23\,\mathrm{Myr}$ are 0.7 and 1.2~kpc, respectively, taken by inspection of the density slice (Fig.~\ref{slice BE}).}
    \label{n SFR BE}
\end{figure*}

\item The bi-modality of $\rhoSFR$ (as seen in Fig.~\ref{n_SFR sim D}) is a clear implication of the jet feedback on the host galaxy, as can be inferred in Fig.~\ref{n SFR BE}. The distinction between the inner ($R<R_\mathrm{jet}$) and outer ($R>R_\mathrm{jet}$) region in terms of $\rhoSFR$ is stronger in Sim.~E, where the jet is inclined and strongly interacts with the gas for a longer time. The mean densities of clouds inside the region of direct interaction are higher than for clouds in the outskirts. The main difference between Sim.~B and E is that more clouds are affected by the jet in Sim.~E than B, showing a different trend of $\rhoSFR$ in the central region. The outer clouds show behaviour similar to the clouds in the `no-jet' simulation, but with a slight reduction in $\rhoSFR$. This reduction is more for Sim.~E than Sim.~B since the coupling between the jet, and the gas is stronger in Sim.~E, and also more clouds are affected by the jet. Interestingly, again, we see that the blue dashed line always lies below the grey line, implying the regions close to the jet showing enhanced SFR have rates lower than what is expected for clouds with corresponding gas density in the `no-jet' simulation as discussed in Sec.~\ref{individual SFR}.

\begin{figure}
    \centering
    \includegraphics[width=\linewidth]{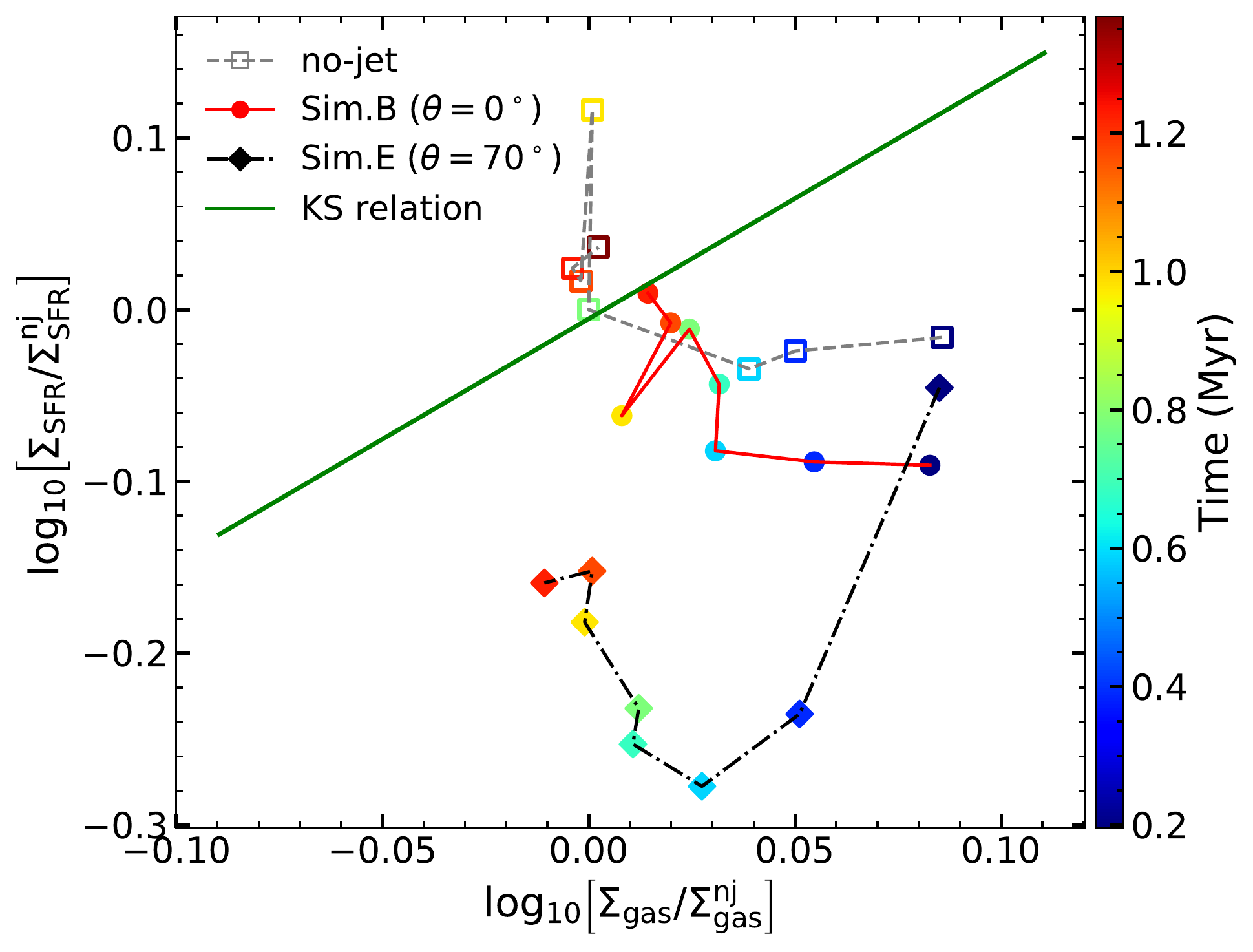}
    \caption{Evolution of the galaxies in the `no-jet' (open square), B (filled circles), and E (diamonds) with time on the normalized KS diagram. The $x$-axis corresponds to the normalized gas surface density. The $y$-axis represents the normalized SFR surface density. We have used the same normalization constant for each variable as in Fig.~\ref{KS sim D}.}
    \label{KS BE}
\end{figure}

\item The evolution of $\Sigma_\mathrm{SFR}$ as a function of $\Sigma_\mathrm{gas}$ with time for Sim.~B and E is presented in Fig.~\ref{KS BE}. The open squares are the galaxy in the `no-jet' simulation at different times indicated in the colourbar. The filled circles and diamonds correspond to Sim.~B and E at different times. We see that Sim.~B starts from a relatively lower SFR surface density ($\Sigma_\mathrm{SFR}$) than the `no-jet' simulation and moves towards the KS line almost steadily after that. The $\Sigma_\mathrm{SFR}$ value at each time is a bit lower than the corresponding `no-jet' simulation, showing mild negative feedback. For Sim.~E, the SFR has a similar value as in the no-jet case, initially. However, when the jet starts to evolve through the disc, the SFR decreases, as seen for Sim.~D (Fig.~\ref{KS sim D}). The SFR is even lower than the corresponding SFR for the `no-jet' and Sim.~B, showing relatively stronger negative feedback. However, the feedback is not strong enough to suppress star formation completely. Instead, after the initial suppression, the SFR increases as the jet fails to sustain the velocity dispersion inside the clouds, and radiative shocks increase the cloud mean density. As a result, the star formation efficiency increases again after the initial decline. However, we note that unlike in other cases, Sim.~E does not return to the KS-line and still has reduced SFR at the end of the simulation, as the jet is still actively interacting with the gas disc.
\end{enumerate}

\section{Discussion}\label{discussion}
In this study, we have applied a sub-grid prescription for star formation in the simulations of the jet-ISM interaction from \citet{Mukherjee_2018}. This is one of the first efforts to estimate the impact of relativistic jets on the host galaxy's ISM, accounting for various different physical parameters such as the cloud density, velocity dispersion, and Mach number\citep[see][for similar such applications to other large-scale simulations]{Dobbs_2013,Ward_2016,Nickerson_2019,Hui_2020}. We find important qualitative differences compared to simulations that adopt a constant star formation efficiency above a fixed density threshold, which allows us to examine new aspects of how radio jets interact with their surrounding gas and affect the evolution of star formation in the gas disc, as the disc passes through different evolutionary stages. However, since we build upon a previous simulation that had adopted higher initial velocity dispersion within the clouds than what is standard for such work, some of the quantitative results, e.g., the virial parameter, should not be interpreted in the same way. We, therefore, refer a detailed analysis of the quantitative results to a future publication. Thus, in this study, we have set up a framework for estimating the SFR by using the physical properties of the ISM as inputs. In the following, we discuss some of the implications of the results discussed earlier.
\begin{enumerate}
    \item \textbf{Comparison with the previous usage of the SFR efficiency factor in studies of jet-feedback:}
    In most of the studies on star formation on the galactic scale, a density threshold is used as the sole criterion for estimating the SFR, with the assumption that a region above some user-specified density ($\rho_\mathrm{th}$) will eventually undergo gravitational collapse and form stars \citep{Kravtsov_2003,Springel_2003,Dubois_2008,gaibler_2012,Bieri_2015,Dugan_2017}. The SFR per unit volume using this method is usually expressed as,
    \begin{equation}\label{SFR density threshold}
        \mathrm{SFR}=\epsilon_\mathrm{SFR}\frac{\rho}{t_\mathrm{ff}}\propto \rho^{3/2},
    \end{equation}
    where $\epsilon_\mathrm{SFR}$ is the star formation efficiency, equivalent to the SFR per freefall time ($\SFRff$) as discussed in Sec.~\ref{no jet simulation}. $\epsilon_\mathrm{SFR}$ is generally considered to be a constant, i.e.,
    \begin{align*}
        \epsilon_\mathrm{SFR} & = \text{some constant value, \,\,\, if } \rho > \rho_\mathrm{th,}  \\
                              & = 0. \hspace{2.7cm} \text{otherwise.}   
    \end{align*}
   The assumption that $\epsilon_\mathrm{SFR}$ is constant for any cloud condition is highly simplified that is not well motivated physically. The efficiency of star formation for a cloud largely depends on different physical processes that can widely vary depending on the local conditions of the cloud. Applying a constant star formation efficiency for gas above a defined density threshold can lead to a significant overestimation of the global SFR. For our simulations, replacing the multi-freefall model with such a simplistic method leads to global SFR values higher by a factor of a few.
   
   Moreover, from Eq.~\eqref{SFR density threshold} we see that the volumetric SFR density ($\rhoSFR$) varies as $n^{1.5}$. Thus all clouds will show a similar scaling of the SFR and gas densities, missing the bi-modal distribution between the inner and outer clouds demonstrated in the top-left panel of Fig.~\ref{n_SFR sim D}. Such a distribution results from the difference in the strength of interaction of the outflow at different radii of the gas disc and the level of induced turbulence in the clouds, i.e., the SFR depends not only on gas density, but also on the turbulent velocity dispersion \citep[see e.g.,][]{Federrath_2012}.
   
   Indeed, a multi-freefall model of star formation, including the effect of the local physical conditions, shows that the feedback from the jet can create a huge contrast in local dynamical quantities as well as the SFR between the clouds at different regions. We show that although the jet causes a reduction in global SFR due to increased velocity dispersion, the clouds near the jet axis experience an enhanced SFR due to the compression. This can not be predicted from the previous models of star formation using a constant efficiency depending on gas density only. 

    \item \textbf{Feedback: Positive or Negative?}
    It is generally thought that the energy injected by the jet suppresses star formation, i.e., has a negative feedback effect by enhancing internal turbulence and heating the gas through shocks, preventing local gravitational collapse. On the other hand, positive feedback models suggest that radiative shocks from relativistic jets can compress the medium locally  \citep{Wagner_2012,Fragile_2004,Wagner_2016,Bieri_2016, Fragile_2017, Mukherjee_2018}, leading to density enhancements and a subsequent increase in the SFR. Here we find that both these feedback mechanisms can exist in a single system, depending on the evolutionary stages of the jet \cite[see][for a review]{Wagner_2016}. 
    
    When the jet is young and confined inside the ISM, the velocity dispersion and thermal energy increase inside the clouds due to the strong interaction between the jet and the gas. Moreover, the radiative shocks also enhance the density. However, the combined effect of $\sigmav$ and $c_s$ in quenching the star formation efficiency dominates over the density enhancement at this stage, resulting in an overall inefficient star formation. At later times, when the jet decouples from the disc and extends to larger scales, the turbulence decays inside the clouds due to lack of continuous driving, and the density-enhanced clouds cool down, making them prone to collapse \citep{Zubovas_2017}. Hence, both positive and negative feedback can occur at different stages of the evolution for a single jet episode. Similar conclusions have been drawn for galaxies with a quasar-driven outflow as in \citet{Cresci_2015,Shin_2019}. The effect of the jet on the SFR may even disappear if the increase in $\sigmav$ and $n$ are such that they compensate each other mutually. In such a scenario, the gas cooling can significantly affect the dependence of the $\SFRff$ on the temperature and breaks the degeneracy between these two compensating quantities. Thus, the division of the jet energy injected into the kinetic and thermal components of the gas is also an important factor for the feedback mechanism.
    
    In conclusion, our results show  that the jets can indeed reduce the global SFR while simultaneously enhancing the star formation activity at certain points of direct interaction. Although the global reduction in the SFR is only by a factor of a few, our results show qualitative trends that a jet can have a widespread negative impact on the SFR, not just at localised regions as previously envisaged. Whether such a reduction in the star-formation activity can indeed lead to quenching of the SFR by more than an order of magnitude as seen in some radio-loud galaxies \citep[e.g.,][]{Nesvadba_2010,Nesvadba_2011,Alatalo_2011,Alatalo_2015} is yet to be demonstrated. Several factors may contribute to weakening the suppression in SFR that we find here: imperfect initial conditions, the absence of a better cooling model at temperatures below $1000~\mathrm{K}$, and limited numerical resolution. Moreover, dynamically coupling the SFR model (used here only in post-processing) to the simulations may further change the evolution and impact of the relativistic jet. We aim to tackle these challenges in future studies.
    
\end{enumerate}

\section{Summary and Conclusion}\label{conclusion}
In this paper we have studied the impact of a relativistic jet on the star formation in a galaxy disc. A subgrid model has been implemented for computing the SFRs inside individual star-forming clouds, identified by a CLUMPFIND routine. We have explored the effect of the jet feedback on different dynamical quantities that regulate the SFR inside individual clouds, e.g., the velocity dispersion, virial parameter, mean density of the clouds, etc. In the following, we summarize the main results of this paper.

\begin{itemize}
    \item The collimated jet mainly affects the surroundings of the jet axis. The outskirts of the disc remain almost intact or get mildly affected by the backflows. The strength and duration of the interaction largely depend on the inclination of the jet with respect to the disc.  
    
    \item Powerful shocks from the jet induce turbulence near the region of direct interaction. This enhances the mean gas density in the clouds through compression and also increases the velocity dispersion within clouds. However, after the jet breaks out of the disc, creating a channel through the ISM, most of the jet energy escapes. As a result, the velocity dispersion inside the dense gas decreases as the turbulence decays without continuous driving from the jet.
    
    \item The interaction with the radio jet generates a bi-modal distribution of the volumetric SFR density ($\rhoSFR$) among the clouds. The dense clouds near the central region, where the density is enhanced due to the compression, exhibit a generally higher SFR than the clouds in the outskirts. However, the efficiency of converting the gas into stars is somewhat lower than what should be expected for clouds with similar density in the `no-jet' simulation. However, when the jet breaks out of the disc, the velocity dispersion decreases inside the clouds, making the clouds more efficient in forming stars, and the bi-modality of $\rhoSFR$ disappears. Hence, the confinement time-scale of the jet inside the disc has important consequences for the star formation activity.
    
    \item We notice a morphological disruption of the distribution of the SFR surface density compared to the `no-jet' case, i.e., a ring-like enhanced star formation region near the central cavity created by the outflow. The clouds near the central region experience the strongest compression by the inflating bubble from the jet, which efficiently compresses the gas to a high density, resulting in an enhanced star formation efficiency when the internal turbulence has decayed to a sufficiently low velocity dispersion. 
    
    \item A single system can go through different evolutionary stages depending on the intensity of the jet-ISM interaction. Initially, when the jet starts to interact with the disc, the velocity dispersion increases, which reduces the efficiency of star formation inside the clouds, resulting in a decline in the global SFR. The jet-driven outflows also ablate and fragment clouds, leading to an overall reduction in SFR, as in conventional models of negative feedback. However, the compression also enhances the mean density of the clouds. At later times, due to the dying turbulence and absence of driving, and also the enhanced density, the SFR efficiency increases. Although the global SFR remains lower than that of the `no-jet' simulation, the SFR of the jetted simulation shows a relative increase compared to the early decline. This is akin to positive feedback, although still inefficient compared to the `no-jet' simulation. Hence, both modes of feedback can occur in a single system depending on the evolutionary stage of the jet. However, we note that the magnitude of the suppression of the SFR is only a factor of a few. Nevertheless, our study provides interesting qualitative first evidence that jet feedback has a potential to lower the SFRs in galaxies through the modification of the turbulence in the ISM rather than by driving gas out of the galaxy. Future simulations directly including star-formation and stellar feedback and conducted at higher spatial resolution so that different phases of the ISM are better resolved are required to quantify the role of AGN jets in affecting the overall star formation rate of its host galaxy in more detail. 
\end{itemize}

\section*{Acknowledgements}
We thank the anonymous referee for their constructive comments, which helped to improve this work. We gratefully acknowledge use of the high performance computing facilities at IUCAA, Pune\footnote{\url{http://hpc.iucaa.in}}. C.~F.~acknowledges funding provided by the Australian Research Council (Future Fellowship FT180100495), and the Australia-Germany Joint Research Cooperation Scheme (UA-DAAD). AYW is supported by JSPS KAKENHI Grant Number 19K03862. The simulations of \citet{Mukherjee_2018}, whose results have been used to investigate the SFR in this work, were undertaken with the assistance of resources and services from the National Computational Infrastructure (NCI, project codes n72 and ek9), which is supported by the Australian Government.

\section*{Data Availability}
No new data were generated in support of this research. The simulations used in this work are available from the corresponding authors upon reasonable request.
\FloatBarrier




\bibliographystyle{mnras}
\bibliography{mainmanuscript} 




\appendix
\section{A short review of turbulence regulated star formation theory}\label{append:SFR theory}
In this section we briefly summarise the theoretical estimates of  star formation rate (SFR) in a turbulent ISM developed by \citet{Krumholz_2005} and later expanded in \citet{Federrath_2012}. The methods have been widely applied to understand the SFR  in numerical simulations of supersonic turbulence \citep{Federrath_2010b,Padoan_2011,Federrath_2012,Padoan_2012} and also observations of molecular clouds \citep{Salim_2015}.

The star-formation rate in a region depends on the relative contributions of the gravitational potential energy and the turbulent kinetic energy. Stars are expected to form at scales where turbulent pressures do not affect density fluctuations. This is considered to occur at the the sonic scale ($\lambda_s$),  where the turbulent velocity dispersion becomes comparable to the local sound speed \citep{Padoan_1995,Vazquez_2003,Krumholz_2005,Kritsuk_2007a,Federrath_2012}.
Assuming the scale dependence of the turbulent velocity dispersion to be $\sigma_v \propto l^{0.5}$ as established by observations \citep{Larson_1981,Solomon_1987,Ossenkopf_2002,Heyer_2009,Roman_Duval_2011} and numerical simulations \citep{Kritsuk_2007a,Schmidt_2009,Federrath_2010,Federrath_2021}, the sonic scale from the above definition is given by
\begin{equation}
c_s = \sigmav \left(\frac{\lambda_s}{L}\right)^{0.5}
\end{equation}

The Jean's length of a region with density $\rho$ and sound speed $c_s$ is 
\begin{equation}
    \lambda_J(\rho)  = \left(\frac{\pi c_s^2}{G \rho}\right)^{1/2}.
\end{equation}
If star formation occurs at scales below the sonic scale, then $\lambda_s \sim \lambda_J(\rho) = \phi_x \lambda_J(\rho)$. The coefficient $\phi_x$, is obtained from fits to the results from numerical simulations, and varies between $\sim 0.1-1$ \citep{Krumholz_2005,Federrath_2012}. Thus when compared to the value of the Jeans length at the mean density ($\rho_0$), we get a natural limit on the density ($\rho_{\rm crit}$) beyond which star-formation is expected to take place \citep{Krumholz_2005,Federrath_2012}:
\begin{align}
    \frac{\lambda_J(\rho_0)}{\lambda_J(\rho)} &= \phi_x \frac{\lambda_J(\rho_0)}{\lambda_s} \propto \left(\frac{\rho_{\rm crit}}{\rho_0}\right)^{1/2} \\
    \rho_{\rm crit} &= \rho_0 \left(\phi_x \frac{\lambda_J(\rho_0)}{\lambda_s }\right)^2
\end{align}
The critical density can be expressed in a more convenient form as \citep{Federrath_2012}:
\begin{equation}\label{a:rho_critical}
    \rho_\mathrm{crit}=\rho_0\left[\left(\frac{\pi^2}{5}\right)\phi_x^2\alpha_\mathrm{vir}\mathcal{M}^2\right].
\end{equation}
Here $\mach$ is the mean rms Mach number. The virial parameter $\alphavir=2E_\mathrm{kin}/E_\mathrm{grav}$ measures the relative strength of the turbulent kinetic energy of the clump to its gravitational potential energy.

The star formation rate in a region will thus depend on the fraction of total mass  beyond the critical density, normalised to a gas in-fall time scale ($t_{\rm in}$). If the probability distribution function (PDF) of the density is given by $p(s)$, with $s = \ln(\rho/\rho_0)$, the SFR will then be
\begin{equation}
    \mathrm{SFR} \propto \int_{s_\mathrm{crit}}^\infty\frac{\rho p(s)}{t_\mathrm{in}} ds.
\end{equation}
 The gas in-fall time is taken to be proportional to the local free fall time as
\begin{equation}
    t_{\rm in} = \phi_t t_{\rm ff}; \quad t_{\rm ff} = \left(\frac{3 \pi}{32 G\rho}\right)^{1/2}.
\end{equation}
The coefficient $\phi_t$ is  found from numerical simulations to be in the range $\phi_t \sim 0.3 - 2$ \citep{Krumholz_2005,Federrath_2012}.

Using the above, one can define the star formation rate per free fall time ($\SFRff$) as \citep{Krumholz_2005,Federrath_2012}:
\begin{equation}\label{a:eq.SFRff}
\mathrm{SFR_{ff}} =\frac{\epsilon}{\phi_t}\int_{s_\mathrm{crit}}^\infty\frac{t_\mathrm{ff}(\rho_0)}{t_\mathrm{ff}(\rho)}\frac{\rho}{\rho_0}p(s)ds .
\end{equation}
The dimensionless quantity $\SFRff$ represents the fraction of total gas mass beyond a critical density that can form stars per gas in-fall time scale. The total star formation rate of a gas clumps with mass $M_c$ and mean density $\rho_0$ is then 
\begin{equation}\label{a:eq.SFR}
    \mbox{SFR} = \frac{M_c}{t_{\rm ff}(\rho_0)} \SFRff
\end{equation}
Here $\epsilon$ denotes the efficiency with which the dense gas is converted to stars. The efficiency at local star forming sites is usually not more than a few percent \citep{Matzner_2000,Alves_2007,Andre_2010,Federrath_2012} due to strong feedback from star forming regions which can lead to significant mass loss through outflows.

From the above discussion, it becomes evident that knowledge of the critical density  of star formation from Eq.~\eqref{a:rho_critical} and the density distribution ($p(s)$) fully characterises the SFR of a clumps. In our simulations of jet-ISM interactions, large scale, dense contiguous gas clumps with densities higher than a threshold ($n > 100 \cc$) clumps and sizes ($L\gtrsim 60$ pc) can be identified. Gas clumps of such sizes and densities are typical of giant molecular clouds \citep{Hughes_2010,Hughes_2013,Miville_2017,Faesi_2018}. The critical density can be trivially evaluated from the fluid parameters of a simulation. However, the true extent of the density PDF of a selected clump is difficult to obtain accurately. Since the clumps are selected by first assuming a density threshold of $n > 100 \cc$, the low density part of the PDF of a given clump remain under-sampled. However, one can extend the theory of supersonic turbulence to describe the properties of gas clumps, as done while developing the analytical theory of SFR in several papers such as \citet{Krumholz_2005,Padoan_2011,Hennebelle_2011,Federrath_2012}, as we describe below.

As discussed in Sec.~\ref{methods}, the density structure of isothermal supersonic turbulence follows a log-normal distribution (Eq.~\ref{eq:density PDF}). The PDF is usually expressed in terms of the density contrast $\tilde{\rho} = \rho/\rho_0$ with respect to the mean density $\rho_0$ as
\begin{equation}\label{a:density pdf}
    p_s(s)=\frac{1}{\sqrt{2\pi\sigma_s^2}}\exp\left[-\frac{(s-s_0)^2}{2\sigma_s^2}\right]\,;\quad \mbox{ with }s = \ln(\tilde{\rho})
\end{equation}
where $s_0$ is the logarithmic mean and $\sigma_s$ the dispersion. The normalisation constraint arising from definition of the mean of the PDF $[\tilde{\rho}] = \int \tilde{\rho} p(s) ds = 1$, relates the mean density and dispersion as \citep{Ostriker_Stone_2001,Li_Klessen_2003,Federrath_2012}
\begin{equation}\label{a:eq.s0}
       s_0=-\frac{1}{2}\sigma_s^2.
 \end{equation}
The logarithmic density dispersion ($\sigma_s$) of the PDF for a hydrodynamic gas is defined by its Mach number as (Sec.~\ref{methods}):
\begin{equation}\label{a:eq.sigma_s_HD}
    \sigma_s^2=\ln\left[1+b^2\mathcal{M}^2\right]
\end{equation}
The parameter $b$ denotes the ratio of energies in the solenoidal and compressive modes of turbulence \citep{Federrath_2010}, which we set to $b=0.5$ in this paper. 
Thus for an assumed lognormal distribution of the density, one can integrate Eq.~\eqref{a:eq.SFRff} to get 
\begin{align}\label{a:eq.SFRff_integrated}
\mathrm{SFR_{ff}} &=\frac{\epsilon}{2\phi_t}\left[1+\mathrm{erf}\left(\frac{\sigma_s^2-s_\mathrm{crit}}{\sqrt{2\sigma_s^2}}\right) \right]\exp\left[\frac{3}{8}\sigma_s^2\right],
\end{align}
Eq.~\eqref{a:eq.SFRff_integrated} together with Eq.~\eqref{a:eq.SFR} gives the SFR of a selected density clump in the simulation. Note that Eq.~\eqref{a:eq.sigma_s_HD} and Eq.~\eqref{a:eq.SFRff_integrated} can be computed using only three macroscopic properties of the cloud viz. the mass weighted 3D velocity dispersion ($\sigmav$), the rms Mach number ($\mach$) and the gravitational potential. Henceforth, our strategy would be to identify dense potentially star forming gas clumps in the inhomogeneous ISM, and evaluate the SFR using Eq.~\eqref{a:eq.SFRff_integrated}.

\section{The FellWalker algorithm for finding clumps}\label{append:Fellwalker}
The FellWalker algorithm for clump detection was first proposed and implemented by \citet{BERRY_2015}. This is a gradient tracing algorithm in which the algorithm propagates from a low-valued pixel to a significant peak in the data value, following an upward direction. The key concept of this algorithm consists of walking along every alternative path of steepest ascent (maximum gradient) from a low-value pixel to reach a peak, each of which is the part of a clump. A 'walk' is defined as a series of steps in which the algorithm begins from a pixel and progresses to a neighbouring pixel of higher value, until a significant peak is reached. A walk can be initiated by any pixel with a data value greater than a user-defined threshold value. A walk begins with such a pixel and moves the algorithm to a higher-valued neighboring pixel until a pixel is identified whose value is higher than all of its neighboring pixel values, which is considered as a local peak. When the algorithm discovers a local summit, it searches a wider area (usually a nine-pixels-wide box) for a higher value pixel. If a pixel of this kind is located in the larger neighborhood, the walk leaps from the previous local peak to this higher pixel and continues as before. Otherwise, the algorithm determines that a significant peak has been reached and assigns a new clump identifier to all pixels traversed during the walk. If a walk comes across a pixel that is already a member of a clump, the walk is terminated, and all pixels visited so far are allocated to the same clump. The clumps found by this methods are called raw clumps. We refer the readers to \citet{BERRY_2015} for detailed technical method for raw clumps identification.

\subsection{Merging raw clumps}\label{merge}
The maximum distance a walk can leap when looking for a higher valued pixel in a wide neighborhood determines the number of significant peaks (raw clumps) identified by the above procedure. This parameter is known as \textit{MaxJump}. When \textit{MaxJump} is set to a higher value, more local peaks are perceived as noise spikes rather than significant peaks, resulting in a decrease in the number of significant peaks. As a result, the peaks are distinguished based on their spatial separation at this stage. This easily illustrates that noise spikes separated by more than MaxJump distance would fragment a clump with a large and flat summit into several small clumps. FellWalker imposes a merging algorithm between the clumps to prevent artificial fragmentation caused by noise spikes. Here, our criterion for merging the raw clumps differs from the one used in the original FellWalker algorithm described in \citet{BERRY_2015}. The original concept behind merging adjacent clumps was to find the highest value on the border between two clumps. Two clumps will be combined into one clump if this value is greater than a user-specified parameter \textit{MinDip} (depending on the data noise level). Here the value of \textit{MinDip} is absolute. The problem with this absolute value is the following. Let us consider that the threshold is low, and we find two clumps such that they have very high peak values with a distinguishable valley between them. However, the highest value along the boundary between these two clumps is higher than the \textit{MinDip} value. So, these two clumps will be merged if we follow the original merging criteria. However, for many statistics, we want these to be two separate clumps. Thus, we should merge two clumps depending on the peak value as well as the dip value. In order to do so, we make the \textit{MinDip} value a variable parameter that depends on the peak values of two neighboring clumps. We introduce a new user-defined parameter ``\textit{per}'' which determines the \textit{MinDip} value for merging two clumps as the percentage of the maximum between the peak values of two neighboring clumps under consideration i.e. 
	\begin{equation}\label{per}
	MinDip=per\times max\text{(peak values of two neighboring clumps)} 
	\end{equation}
Thus, two clumps will be merged if the highest point on the boundary between the two clumps is higher than a certain percentage of the clumps' maximum peak values.

After the merging of the clumps, we finally impose the minimum size criteria of a clump. If the number of the cells in a clump is greater than some user-specified value, then that clump will be considered for analysis. Otherwise, we drop the clump. In Table.~\ref{parameter FellWalker}, we enlist all the parameter values for identification of clumps from the simulations data. 

\begin{figure*}
\centerline{
\def\arraystretch{1.0}
\setlength{\tabcolsep}{0.0pt}
\begin{tabular}{lcr}
  \includegraphics[width=0.33\linewidth]{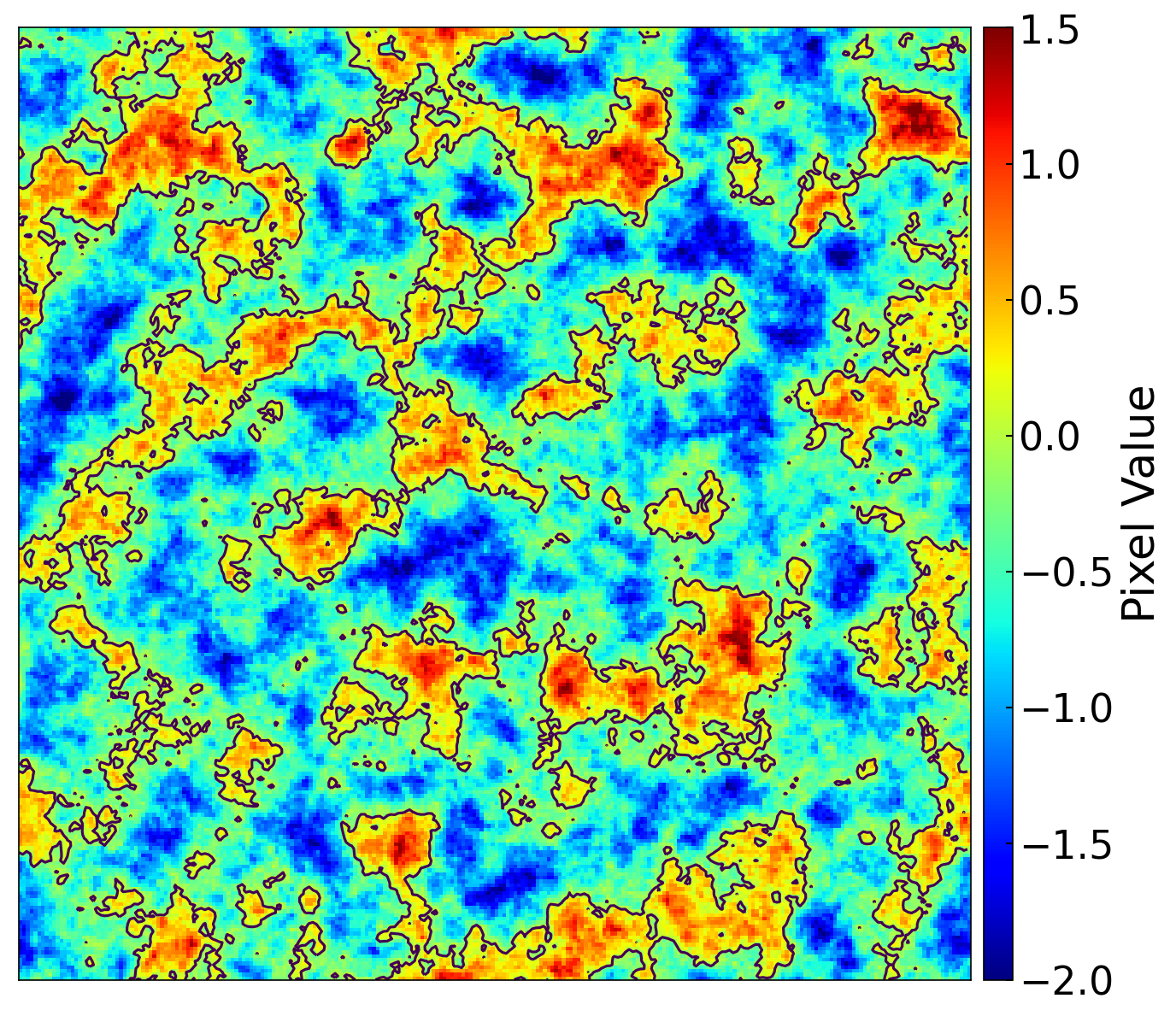} &
  \includegraphics[width=0.285\linewidth]{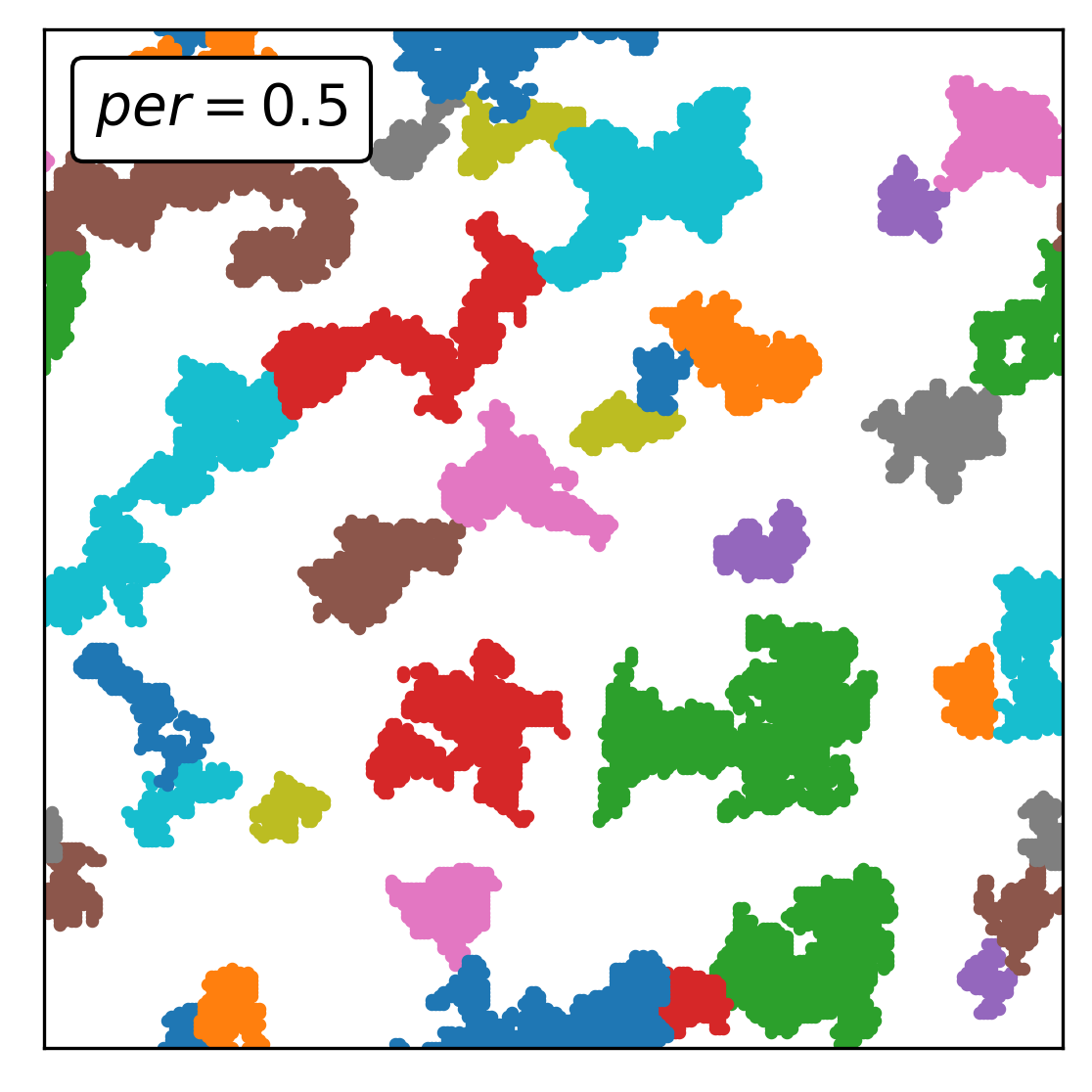} &
  \includegraphics[width=0.285\linewidth]{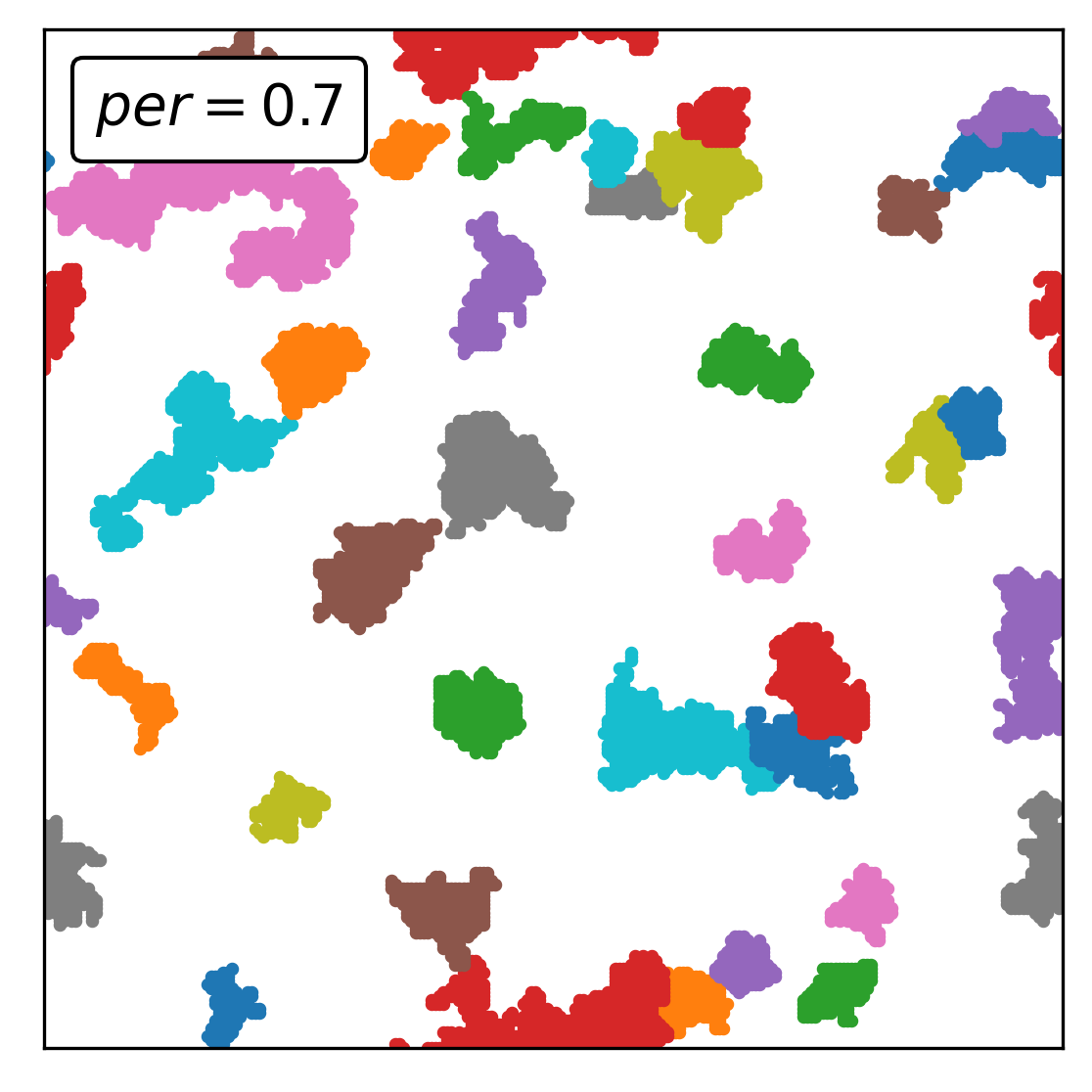}
\end{tabular}}
  \caption{A 2D demonstration of the clumps after merging identified by the FellWalker algorithm \textit{Left:} An artificial $256\times$ 256 log-normal field in 2D. The black lines are the contours at value 1 used as the threshold for clump identification. \textit{Middle:} Identified clumps by the algorithm for \text{per} value of $0.5$. \textit{Right:} Same as the middle panel but for a per value of $0.7$.}
  \label{clumps}
\end{figure*}

\begin{table}
 \caption{Parameter values used in this study for identifying clumps in the simulations.}
 \label{parameter FellWalker}
 \begin{tabular*}{\columnwidth}{c @{\extracolsep{\fill}} c}
  \hline
  Parameter & Value \\
  \hline
  Threshold value & 100\\
  Larger neighbourhood width & 9 \\
  \textit{MaxJump} & 7\\
  \textit{per} & 0.7 \\
  Minimum cells per clump & 1000 \\
  \hline
 \end{tabular*}
\end{table}

In Fig.~\ref{clumps}, We present a 2D demonstration of the clumps found by the FellWalker algorithm after merging with a density threshold of 1 and \textit{MaxJump} of 7. The left panel is an artificial log-normal field of size $256\times 256$. The black lines are the contour level at threshold. The left and right panel show the clumps identified by the algorithm with a \textit{per} value of 0.5 and 0.7, respectively. We consider clumps those have total pixels greater than 100.

\section{Poisson Solver}\label{append:Poisson}
Solving Poisson's equation is ubiquitous in any physical system where gravity is involved. For a given source distribution $\rho$, if $\phi$ is the potential due to the source inside a region of interest, then
\begin{equation}\label{elliptic}
	\nabla^2\phi=\rho
\end{equation}
In this study, we use the \textit{successive overrelaxation} (SOR) method combined with the \textit{Chebyshev acceleration} to solve the Poisson's equation in the computational grid \citep[see][for detailed algorithm]{Press_1992}.

\subsection{Boundary Condition}
The analytical calculation of gravitational potential assumes that the potential becomes zero at a very large distance from the source distribution. However, it is not feasible to capture the spatial infinity in a finite computational domain. Thus, we need to approximate the potential on the domain boundary. The easiest choice is to calculate the potential by direct summation treating the grid cells as point sources, i.e.,
\begin{equation}
    \phi_{m,n,p}=-G\sum_{i,j,k}\frac{\rho_{i,j,k}\,\Delta V_{i,j,k}}{|\boldsymbol{r}_{i,j,k}-\boldsymbol{r}_{m,n,p}|}
\end{equation}
where, $\phi_{m,n,p}$ is the potential value at boundary cell of index $(m,n,p)$ and  $(i,j,k)$ are the cell indices inside the domain. $\Delta V_{i,j,k}$ is the volume of $(i,j,k)^\mathrm{th}$ cell. However, it requires $\mathcal{O}(N^5)$ number of operation to update the boundaries of a domain of size $N^3$, which is not efficient for a moderately large grid. Thus the next natural choice is to find the potential by multipole expansion \citep{Muller_1995,jackson_classical_1999,Katz_2016}. The monopole approximation is the simplest and applied only for a spherical distribution of mass. However, we can approximate a non-spherical distribution by including higher moments. The gravitational potential in terms of the spherical harmonics is given by,
\begin{equation}\label{phi_original}
    \phi(\boldsymbol{r})=-4\pi G\sum_{l=0}^{\infty}\sum_{m=-l}^{l}\frac{1}{2l+1}\int \rho(\boldsymbol{r'})Y_{lm}(\theta,\phi)Y^*_{lm}(\theta',\phi')\frac{r^l_{<}}{r^{l+1}_{>}}d^3\boldsymbol{r}'
\end{equation}
where $\theta$ and $\phi$ are the azimuthal and polar angle, respectively. $r$ is the radial distance and 
\begin{gather*}
    r_<\equiv \mathrm{min}(r,r')\\
    r_>\equiv \mathrm{max}(r,r')
\end{gather*}
Following \citet{Muller_1995} and \citet{Katz_2016} and separating the inner ($r'<r$) and outer ($r'>r$) contributions for even ($\cos$) and odd ($\sin$) terms, one can easily obtain the potential any point ($\boldsymbol{r}$), which is given by, 
\begin{multline}\label{potential full}
    \phi(\boldsymbol{r})=-G\sum_{l=0}^\infty P_{l0}(\cos\theta)\left[r^lQ_{l0}^\mathrm{eo}(r)+\frac{1}{r^{l+1}}Q_{l0}^\mathrm{ei}(r)\right]-\\
    2G\sum_{l=1}^\infty\sum_{m=1}^l P_{lm}(\cos\theta)\Bigg[ (r^l\cos{m\phi}Q_{lm}^\mathrm{eo}(r)+(r^l\sin{m\phi})Q_{lm}^\mathrm{oo}(r)+\\
    \frac{\cos{m\phi}}{r^{l+1}}Q_{lm}^\mathrm{ei}(r)+\frac{\sin{m\phi}}{r^{l+1}}Q_{lm}^\mathrm{oi}(r)\Bigg],
\end{multline}
where $P_{lm}(x)$ are the associated Legendre polynomials. $Q_{lm}$ are the even (e)/ odd (o), inner (i)/ outer (o) moments \citep[see][for detail expressions]{Katz_2016}. 

One can calculate the gravitational potential at any point of the computational domain using Eq.~\eqref{potential full}. However, if the boundary points $(\boldsymbol{r})$ are far away from source distribution $(\boldsymbol{r}')$, then outside contribution will be zero, i.e., $Q_{lm}^\mathrm{eo}$ and $Q_{lm}^\mathrm{oo}$ will be zero. So, we can approximate the potential at the boundary as,
\begin{multline} \label{phi_final}
    \phi(\boldsymbol{r})\approx -G\sum_{l=0}^{l_\mathrm{max}}\frac{Q_{l0}^\mathrm{ei}(r)}{r^{l+1}}P_{l0}(\cos{\theta})\\
    -2G\sum_{l=1}^{l_\mathrm{max}}\sum_{m=1}^l \left[Q_{lm}^\mathrm{ei}(r)\cos{m\phi} + Q_{lm}^\mathrm{oi}(r)\sin{m\phi}\right]\frac{P_{lm}(\cos{\theta})}{r^{l+1}}.
\end{multline}
For practical purposes, we limit the number of multipole moments by restricting $l$ to some finite value $l_\mathrm{max}$ and all of the moments are calculated in the centre of mass frame of the source distribution, which ensures the potential will be dominated by the lower order moments. A similar formalism has also been implemented in the \textsc{Flash} code \citep{Fryxell_2000} and \textsc{Castro} code \citep{Almgren_2010}. 

In this study, for solving Poisson's equation for each cloud, we calculate the boundary term using Eq.~\eqref{phi_final}. To ensure that the density distribution is well inside the domain boundary, we first calculate the maximum number of cells along each axis for a clump and add that number of empty cells at both the boundaries along that axis, i.e., if the maximum extent of a clump is $(N_x,N_y,N_z)$, then the grid size for solving the Poisson's equation is $(3N_x,3N_y,3N_z)$. For this configuration, we find that $l_\mathrm{max}$ value of 5 gives a good approximation of the potential at the domain boundary.  

\section{Effect of the choice of the parameters for defining clumps}
\begin{figure}
    \centering
    \includegraphics[width=\linewidth]{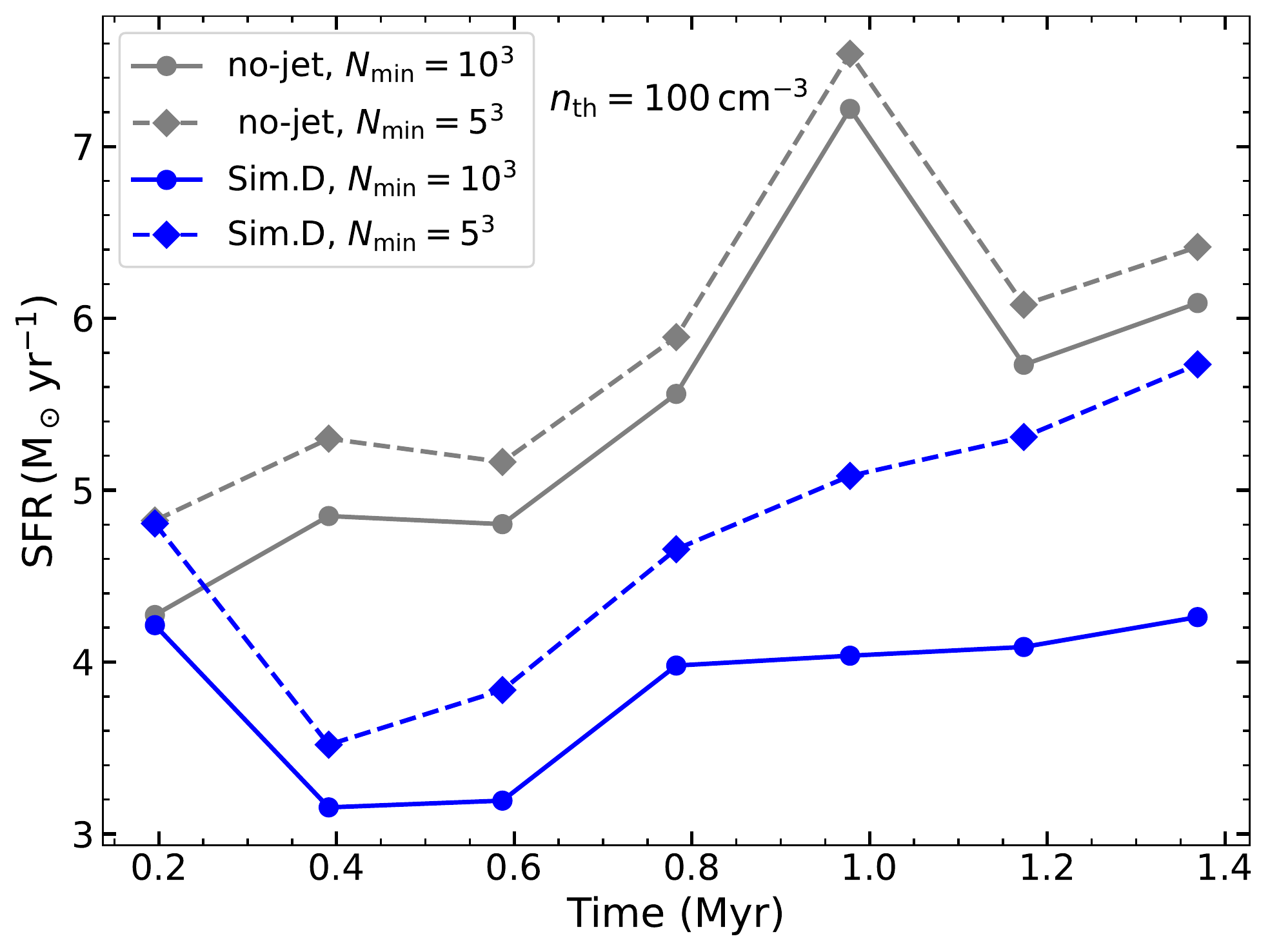}
    \caption{The dependence of the global SFR value on the minimum number of cells ($N_\mathrm{min}$) for defining a clump for `no-jet' simulation (grey) and Sim.~D (blue). The solid and dashed lines correspond to the $N_\mathrm{min}$ value of $10^3$ and $5^3$ for each simulation. The density threshold of $n_\mathrm{th}=100\,\mathrm{cm^{-3}}$ is used for this demonstration.}
    \label{diff resolution}
\end{figure}
\begin{figure}
    \centering
    \includegraphics[width=\linewidth]{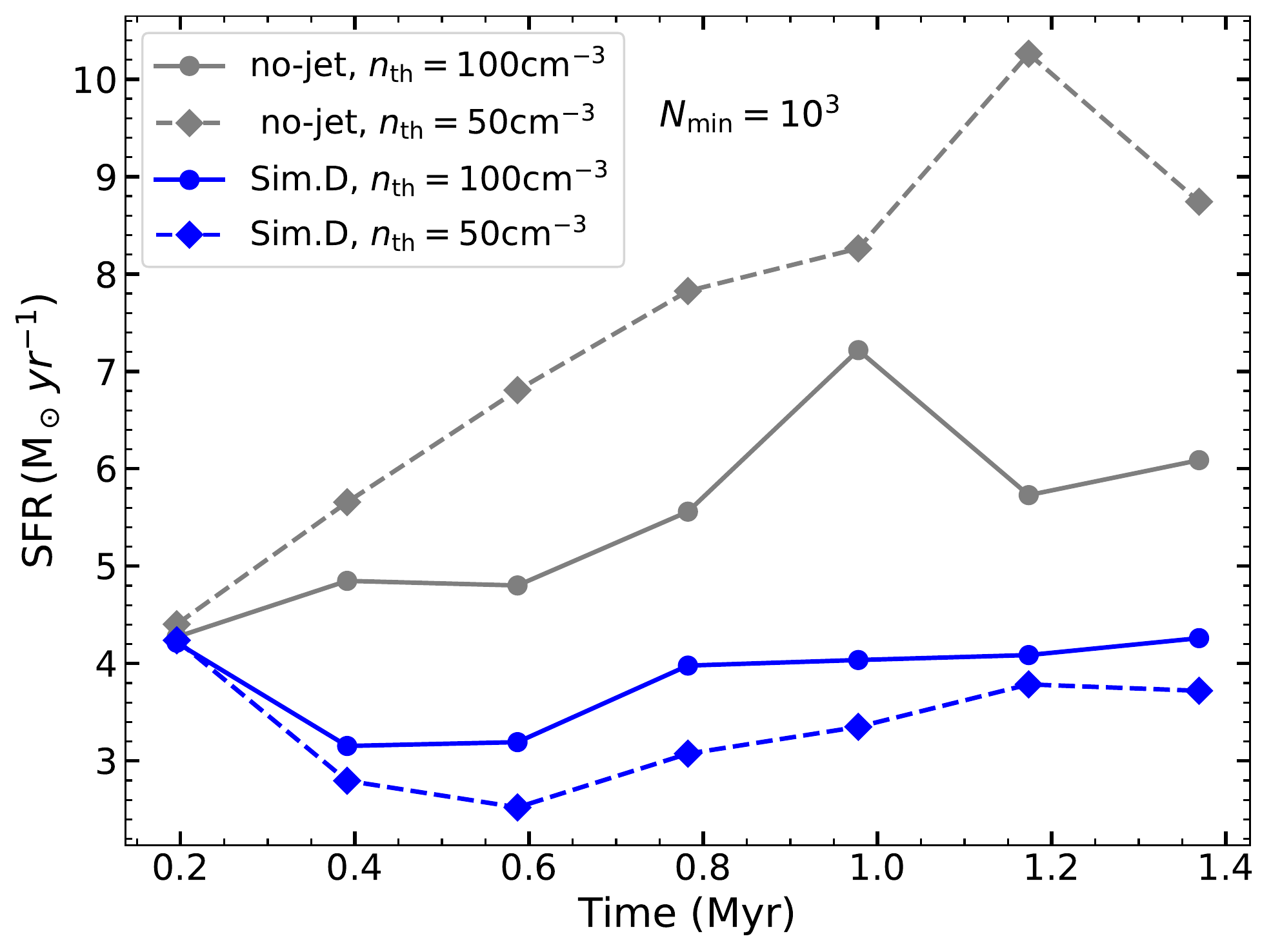}
    \caption{The effect of different density thresholds ($n_\mathrm{th}$) on the evolution of the global SFR. The solid and dashed lines correspond to $n_\mathrm{th}$ value of $100$ and $50\,\mathrm{cm^{-3}}$ respectively. The grey and blue colours represent the results for the `no-jet' simulation and Sim.~D. We consider a $N_\mathrm{min}$ value of $10^3$ for this study.}
    \label{diff threshold}
\end{figure}

In this section, we present the effect of the choice of the minimum number of cells ($N_\mathrm{min}$) and the density threshold ($n_\mathrm{th}$) for defining a clump. In Fig.~\ref{diff resolution}, we show effect of $N_\mathrm{min}$ on the evolution of the global SFR for a $N_\mathrm{min}$ value of $10^3$ (solid) and $5^3$ (dashed). The grey and blue lines correspond to the `no-jet' simulation and Sim.~D. We notice that the SFR values calculated using the $N_\mathrm{min}$ value of $5^3$ have increased by 7\% and 22\% from their values using $N_\mathrm{min}=10^3$ for the no-jet simulation and Sim.~D respectively. This increase in SFR value is due to the resolution of the clouds. By choosing a lower value of $N_\mathrm{min}$, we are unable to compute the velocity dispersion and the Mach number properly, which results in an artificial enhancement of the star-formation rate due to the reduction in the computed value of the velocity dispersion. We would also want to resolve the cloud structure adequately in our simulations. Thus, the criteria of $10^3$ cells inside a cloud give the proper sampling of the structures to compute the local dynamical quantities such as the velocity dispersion, Mach number, etc.

We demonstrate the effect of choosing the density threshold ($n_\mathrm{th}$) on the overall evolution of the global SFR for the `no-jet' simulation (grey) and Sim.~D (blue) in Fig.~\ref{diff threshold}. The solid and dashed lines correspond to the values of $n_\mathrm{th}=100$ and $50\,\mathrm{cm^{-3}}$. We notice completely different behaviour of the change in SFR for the `no-jet' and jetted simulations while changing the $n_\mathrm{th}$ value from $100\,\mathrm{cm^{-3}}$ to $50\,\mathrm{cm^{-3}}$. The global SFR value increases by 34\% for the no-jet simulation, whereas it decreases by 13\% for Sim.~D. A similar reduction of the SFR was checked for other jetted simulations also, and the change was found to be 12\% and 21\% for Sim.~B and Sim.~D, respectively. From this computation, it is clear that there is a fundamental difference between the `no-jet' and jetted simulations. By lowering the density threshold, we consider more low-density gas for a cloud. For the jetted simulations, such low density gas has an overall higher velocity dispersion and higher temperature due to the energy injection from the jet into the ISM. However, for the `no-jet' simulations, the velocity dispersion is almost the same as there is no strong external driver. Thus, by lowering the density threshold in the `no-jet' simulations, we include more mass from the low-density regions, which increases the SFR. On the other hand, for the jetted simulations, such low-density gas does not contribute to the mass budget as can be seen from density PDF in Fig.~\ref{density pdf sim D} but increases the overall velocity dispersion, which reduces the global SFR.

We note here that our choice of $n_\mathrm{th}=100\,\mathrm{cm^{-3}}$ for this study is motivated mainly by two criteria. First, the dense molecular star-forming clouds have mean densities greater than $100\,\mathrm{cm^{-3}}$ \citep{Hughes_2010,Hughes_2013,Miville_2017,Faesi_2018}. Second, from the evolution of the density PDF, we notice that as the clouds get impacted by the jet, the outer layers get ablated and disperse into the low-density region. Such gas can not be considered as star-forming materials. Thus, we believe our choice of $n_\mathrm{th}$ of $100\,\mathrm{cm^{-3}}$ is a better choice given the fact that we see molecular clouds with $n>100\,\mathrm{cm^{-3}}$ where star formation takes place.


\bsp	
\label{lastpage}
\end{document}